\documentclass[lettersize,journal]{IEEEtran}
\usepackage{amsmath,amsfonts}
\usepackage{algorithmic}
\usepackage{algorithm}
\usepackage{array}
\usepackage[caption=false,font=normalsize,labelfont=sf,textfont=sf]{subfig}
\usepackage{textcomp}
\usepackage{stfloats}
\usepackage{url}
\usepackage{verbatim}
\usepackage{graphicx}
\usepackage{cite}
\usepackage{amssymb}
\usepackage{hyperref}
\usepackage{acronym}
\usepackage[table]{xcolor}
\usepackage{color,soul}
\usepackage{xcolor}
\usepackage{tikz}
\usepackage{textcomp}
\usepackage{lipsum}

\acrodef{RIS}{Reconfigurable Intelligent Surface}
\acrodef{LIS}{Large Intelligent Surface}
\acrodef{EM}{Electromagnetic}
\acrodef{HSF}{HyperSurface}
\acrodef{MS}{Metasurface}
\acrodef{QoS}{Quality of Service}
\acrodef{LoS}{Line of Sight}
\acrodef{NLoS}{Non-Line of Sight}
\acrodef{mmWave}{millimeter Wave}
\acrodef{3GPP}{3rd Generation Partnership Project}

\begin{document}

\markboth{IEEE Transactions on Nanotechnology -- DOI: 10.1109/TNANO.2022.3195116}{IEEE Transactions on Nanotechnology -- DOI: 10.1109/TNANO.2022.3195116}

\title{On the Enabling of Multi-receiver Communications with Reconfigurable Intelligent Surfaces}

\author{Hamidreza~Taghvaee\href{https://orcid.org/0000-0001-8732-6086}{\includegraphics[scale=0.5]{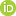}},
Akshay~Jain, 
Sergi~Abadal,
Gabriele~Gradoni,
Eduard~Alarc\'{o}n,
and Albert~Cabellos-Aparicio
\thanks{H. Taghvaee and G. Gradoni are with the George Green Institute for Electromagnetics Research, Department of Electrical and Electronics Engineering,
University of Nottingham, Nottingham NG7 2RD, United Kingdom (e-mail: hamidreza.taghvaee@nottingham.ac.uk)}

\thanks{A. Jain is with the Telecommunications Engineering Dept., Neutroon Technologies S.L., Barcelona, Spain} 
\thanks{S. Abadal, E. Alarc\'{o}n and A. Cabellos-Aparicio are with the NaNoNetworking Center in Catalonia (N3Cat), Universitat Polit\`{e}cnica de Catalunya, 08034 Barcelona, Spain}
\thanks{\textcolor{red}{\textcopyright \quad 2022 IEEE. Personal use of this material is permitted, but republication/redistribution requires IEEE permission. Refer to IEEE Copyright and
Publication Rights for more details.\\
Digital Object Identifier (DOI): 10.1109/TNANO.2022.3195116}}}



\maketitle

\begin{abstract}
The reconfigurable intelligent surface is a promising technology for the manipulation and control of wireless electromagnetic signals. In particular, it has the potential to provide significant performance improvements for wireless networks. However, to do so, a proper reconfiguration of the reflection coefficients of unit cells is required, which often leads to complex and expensive devices. To amortize the cost, one may share the system resources among multiple transmitters and receivers. In this paper, we propose an efficient reconfiguration technique providing control over multiple beams independently. Compared to time-consuming optimization techniques, the proposed strategy utilizes an analytical method to configure the surface for multi-beam radiation. This method is easy to implement, effective and efficient since it only requires phase reconfiguration. We analyze the performance for indoor and outdoor scenarios, given the broadcast mode of operation. The aforesaid scenarios encompass some of the most challenging scenarios that wireless networks encounter. We show that our proposed technique provisions sufficient improvements in the observed channel capacity when the receivers are close to the surface in the indoor office environment scenario. Further, we report a considerable increase in the system throughput given the outdoor environment.
\end{abstract}


\section{Introduction}
\label{sec:introduction}
\IEEEPARstart{W}{ireless} data rates have been increasing exponentially and continue to double every 18 months \cite{1309810}. To keep up with such an explosion in data rate requirements, technologies that can provide faster, sustainable, and safer communications are essential. Further, congestion of the overcrowded \ac{EM} spectrum limits the ever-increasing demand for faster data rates \cite{AKYILDIZ20103}. This has motivated the migration of wireless networks toward utilization of carrier waves with higher frequencies. The \ac{mmWave} spectrum can offer larger bandwidth and higher bit rates \cite{9737695}. However, mmWaves are compounded by certain well-known issues. High propagation losses and refraction \cite{1491267} combined with the challenge of high power transmitters \cite{hunukumbure2018mmwave}, severely restrict the communication range of mmWave based networks. So, any object can block the \ac{LoS} and this renders \ac{NLoS} communication as a very challenging proposition.

Thus, in order to achieve intelligent, sustainable, and virtual \ac{LoS} communication links, wireless networks have been gradually shifting towards the software-defined paradigm in which all the elements of the network can be controlled via programming. The wireless channel, however, has traditionally remained a non-maneuverable quantity. With the advent of the \ac{RIS} \cite{6206517}, also referred to as \ac{LIS} \cite{8288263}, there has been a fundamental shift towards handling wireless channels, wherein they can now be controlled within the design loop of wireless networks. The explosion of RIS has been in many works that propose to apply it in wireless network \cite{202000783,DAJER2021,9424177,RISE-6G2021} which is a clear testament to the potential impact of the \ac{RIS} concept.

One possible way to realize the \ac{RIS} paradigm is grounded on the powerful \ac{EM} control delivered by the \ac{MS} concept. \acp{MS} are thin layer structures composed of a matrix of sub-wavelength resonators known as \emph{unit cells}. These building blocks allow to manipulate the effective permittivity $\epsilon$ and permeability $\mu$ of the medium \cite{https://doi.org/10.1002/smtd.201600064, Li2018, Chen2016}. With this feature, \ac{EM} characteristics of the reflected wave can be engineered. Such control has been the object of several studies proposing novel absorbers \cite{Taghvaee2014,Taghvaee2017a}, retro-reflectors \cite{Arbabi2017}, optical mixers \cite{Liu2018a}, focusing \cite{PhysRevB.104.235409} and nonlinear devices \cite{Taghvaee2017}. 

Recent works have proven that the behavior of \acp{MS} can be tuned during and after deployment. This is achieved by introducing tunable or switchable electronic components \cite{doi:10.1002/mop.32164} within the \ac{MS} and adding appropriate means of control to achieve (re)programmability. Furthermore, there have been proposals to embed intelligence \cite{s21082765} within the \ac{MS} to make it self-adaptive \cite{Ma2019}, inter-connectable \cite{8788546} or multiband \cite{9737695}.

The overall functionality is derived from the aggregated response of all unit cells, which are tuned individually. Concretely, to realize a particular function (e.g., beam steering), very specific amplitude and phase profiles need to be applied to the impinging wave \cite{PhysRevApplied.11.044024,8745693}. The transition from static to intelligent programmable \acp{MS} indeed promises diverse applications in telecommunication. However, to do so \acp{RIS} need to:
\begin{itemize}
    \item Integrate tuning and control elements on a per-cell basis
    \item Include electronic circuits to implement intelligence within the device
    \item Modify the reflected wavefront with subwavelength \ac{EM} interaction
\end{itemize}

\begin{figure}[!t]
\centering
\vspace{-0.5cm}
\includegraphics[trim={6cm 0 4cm 1cm},clip,width=1\columnwidth]{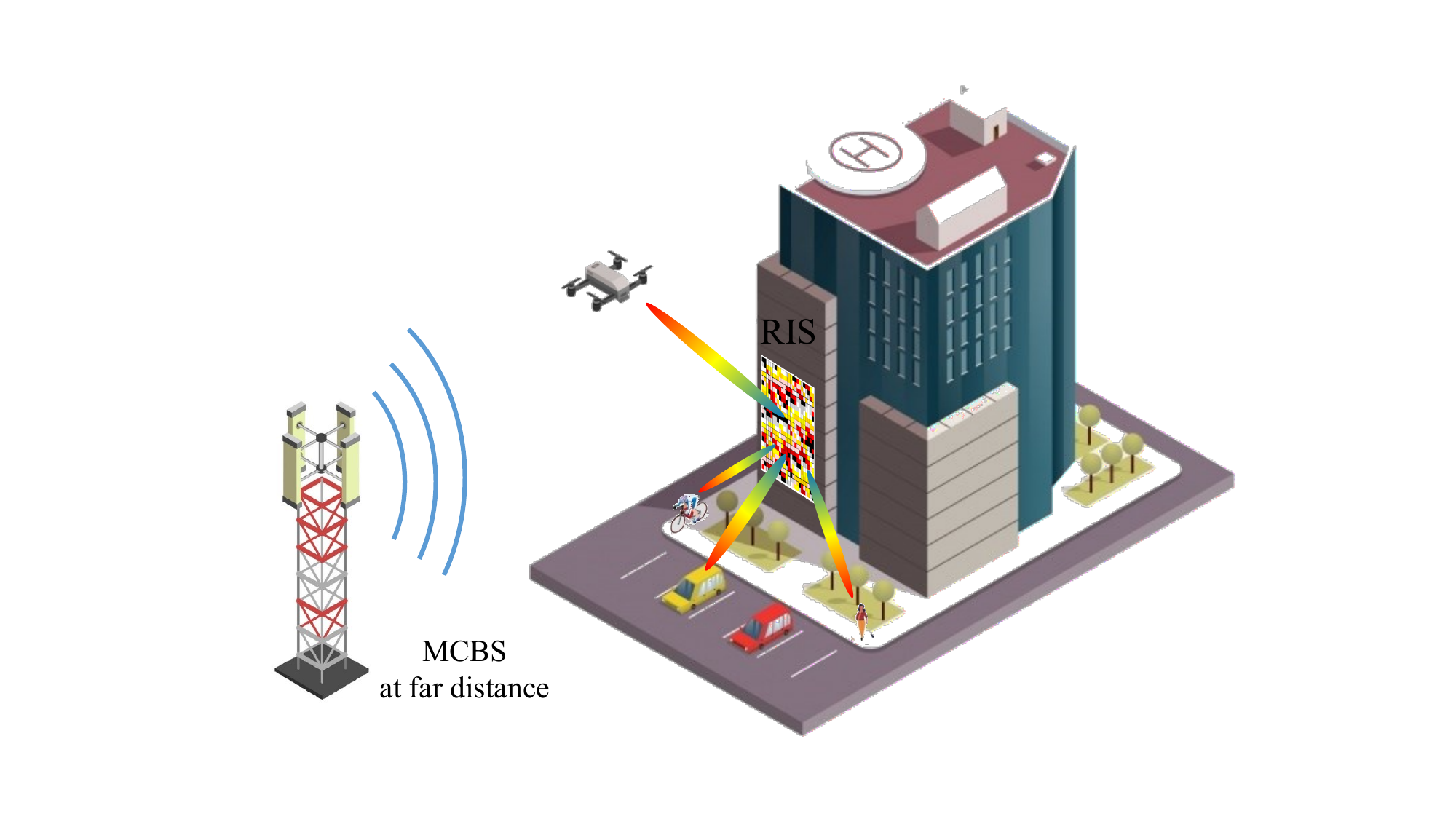}
\vspace{-1.4cm}
\caption{A macro cell base station is in the far field of the location of the users. Since reconfigurable intelligent surfaces are passive devices (in terms of radiation transmission), we are allowed to deploy them in a close range of the users. In addition, multi-beam forming is engineered to serve multiple mobile users.}
\vspace{-.4cm}
\label{fig:ga}
\end{figure}

Such complexity can often lead to uneconomical designs and fabrication processes, which is an obstacle towards commercializing the applications within 5G networks, such as for Vehicle-to-everything communications \cite{V2X,9593176}. One way to justify the costs for utilizing RISs in use-cases as pervasive as V2X (which is a form of multi-receiver communication via a single transmitter) is to optimize their operation. To this end, we note that \acp{MS} can actually perform multiple functions concurrently \cite{9171580}, so one design can serve several purposes. Consequently, multi-receiver communication scenarios present a very compelling use case. 

In a multi-receiver communication scenario, the broadcast station should adequately radiate \ac{EM} waves toward the location of the receivers. A wide beam radiation pattern can provide such a requirement. However, a wide beam is detrimental as it radiates energy over a huge space. This strategy is not feasible for \ac{mmWave} spectrum due to the high propagation losses and blocking effects. The proper solution is to engineer the radiation pattern with respect to the location of the receivers. Hence, independent control on the multiple beams is required. Figure \ref{fig:ga} shows an urban scenario in which the environment is equipped with \ac{RIS} to provide communication services for multiple receivers. 

One way to engineer a multi-beam radiation pattern is to control both amplitude and phase (amp/phs) reflection of the unit cells \cite{8753713}. The amp/phase configuration of a RIS can be derived by calculating the inverse Fourier transform of any required reflected wavefront \cite{Karimipour2019}, but since amplitude reconfiguration increase the overall loss, this is not an efficient approach. Another solution is to switch between the users in the time domain, i.e., time division multiplexing (TDM). However, satisfying the 5G key performance indicators (KPIs) for latency renders the TDM approach inefficient. Additionally, dividing \ac{MS} area i.e., space division multiplexing (SDM), to engineer the wavefront for multiple beam objective requires a very large \ac{MS}. We provide further discussions on the state-of-the-art schemes in Section \ref{sec:back}.

We build this paper based on previous works \cite{9769001,Taghvaee_2021} and we extend it by taking cognizance of all of the above challenges and requirements towards adapting \acp{RIS} for multi-receiver communication environments, we introduce an analytical model to aggregate multi-receiver reconfiguration. Unlike previous methods that require amplitude reflection control as well as phase reflection control of the unit cells \cite{DingChen}, our proposed strategy requires phase reflection reconfiguration only. With this approach the MS realizes multiple-beam radiation pattern with independent control of the beams. Based on realistic system parameters, we then evaluate the performance of the proposed framework by analyzing the throughput for indoor and outdoor scenarios. Given the channel is rank one, this link only supports single data stream to all the users (i.e., broadcast mode). Note that, the broadcast scenario also entails the multicast scenario, which can be utilized by the radio source to communicate with multiple receivers at the same time. We compare our results to the baseline system and show that by taking advantage of the \acp{RIS}, a considerable increase in the overall system throughput can be experienced.     

The organization of the paper is as follows: In Section \ref{sec:back}, we review the latest works in multi-user communications. Section \ref{sec:Model} describes proposed technique to reconfigure the \ac{MS} for multi-beam radiation pattern. Section \ref{sec:scen} describes the indoor, outdoor, and broadcast scenarios on which we evaluate our system. In Section \ref{sec:sysmod} the system model is introduced. Section \ref{sec:eval} presents the performance evaluation and Section \ref{sec:conclusions} concludes the paper.

\section{Background}
\label{sec:back}
\subsection{Time division multiplexing}
\label{sec:tdm}
\begin{figure}[!t]
\centering
\includegraphics[trim={2cm 2.5cm 0cm 2cm},clip,width=1\columnwidth]{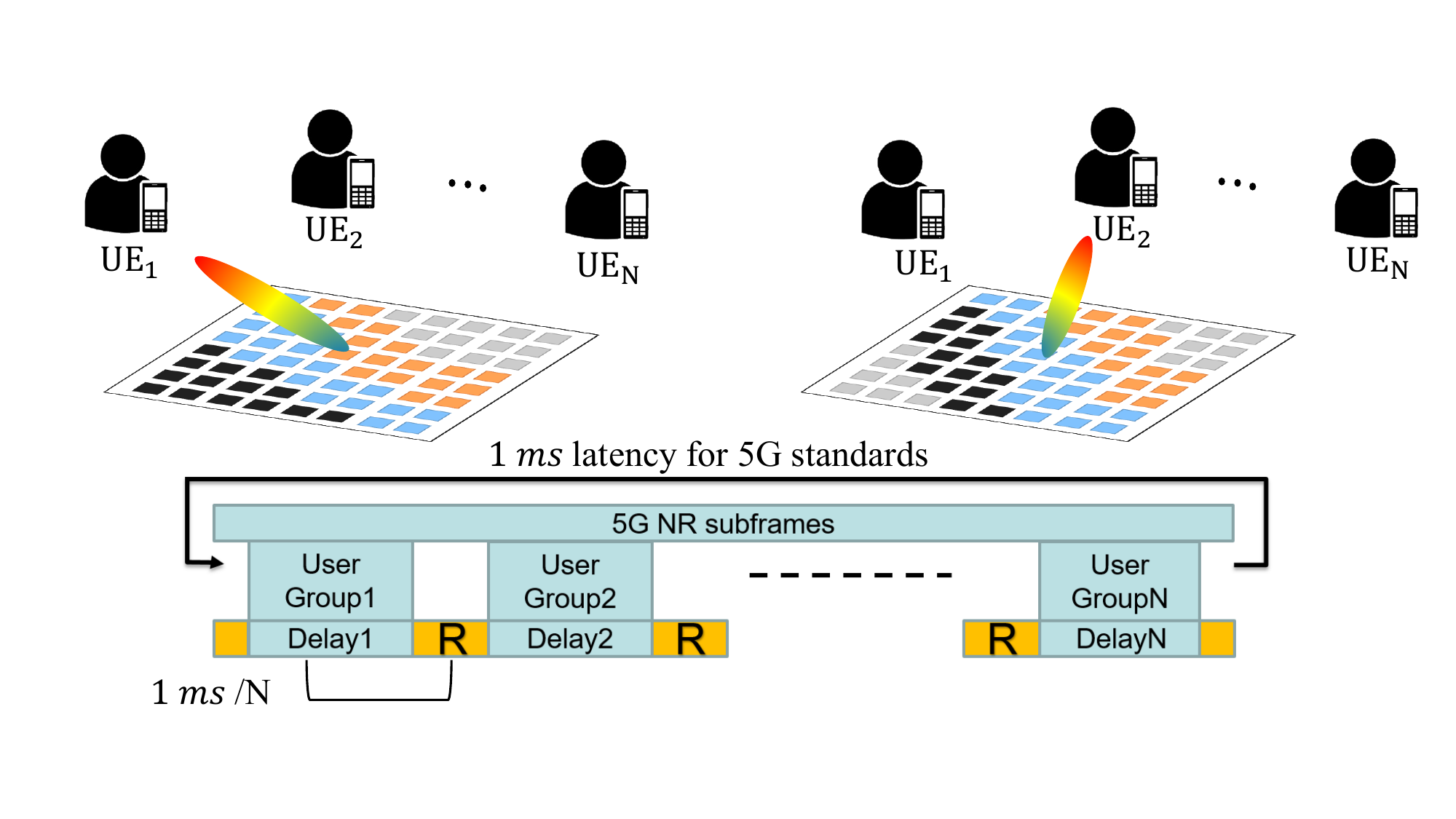}
\caption{Time division multiplexing, switches the beam between the users in time domain. The latency of the link is restricted by the reconfiguration speed and user delay.}
\label{fig:ts}
\end{figure}

TDM allocates the communication link to multiple receivers in separate time slots \cite{Froehlich1991}. Time is divided into several recurrent blocks of fixed length, one for each user. In terms of \ac{MS}, TDM refers to time domain reconfiguration which provides a shared communication link that switches between users. In theory, this technique can provide adaptive multi-channel communication by space-time shared aperture \cite{9220095} with great performance. However, this is not a trivial mechanism, and realizing a TDM \ac{MS} comes with a major challenge. In 5G, the corresponding end-to-end latency as low as 1 $ms$ needs to be met with reliability as high as 99.99$\%$ \cite{7504504}. Tracking a moving receiver requires reconfiguration of the \ac{MS} to sustain the communication link and the reconfiguration speed affects the latency. This might not be a serious problem in point-to-point scenarios but in the multi-receiver case, the reconfiguration cycle is multiplied by the number of receivers. A TDM \ac{MS} switches the link between the receivers in the time domain and the reconfiguration speed of the \ac{MS} will have to be extremely fast to rearrange the link with an acceptable delay. The reconfiguration delay is the time it takes to reprogram the \ac{MS} to serve the specific receiver group (see Fig \ref{fig:ts}). 
\begin{equation}
SL=N\times(UGD+R)
\label{eq:1}
\end{equation}
where $SL$ is the length of the subframe, $UGD$ is the user group delay, $N$ is the number of users and $R$ is the reconfiguration speed. As an example, consider that the maximum length of a single subframe for the 5G New Radio (NR) is 1 $ms$ \cite{7945855}. Further, let us assume that we have $N=10$ groups of users, wherein each group, a user is served in a given subframe. Hence, it is essential that the \ac{MS} reconfiguration is completed in a time that is on the scale of a few microseconds, which is a real challenge. This seems unrealistic for beyond 5G or 6G networks KPIs.

\subsection{Space division multiplexing}
\label{sec:sdm}
\begin{figure}[!h]
\centering
\vspace{-.3cm}
\includegraphics[trim={5cm 0cm 6cm 0cm},clip,width=1\columnwidth]{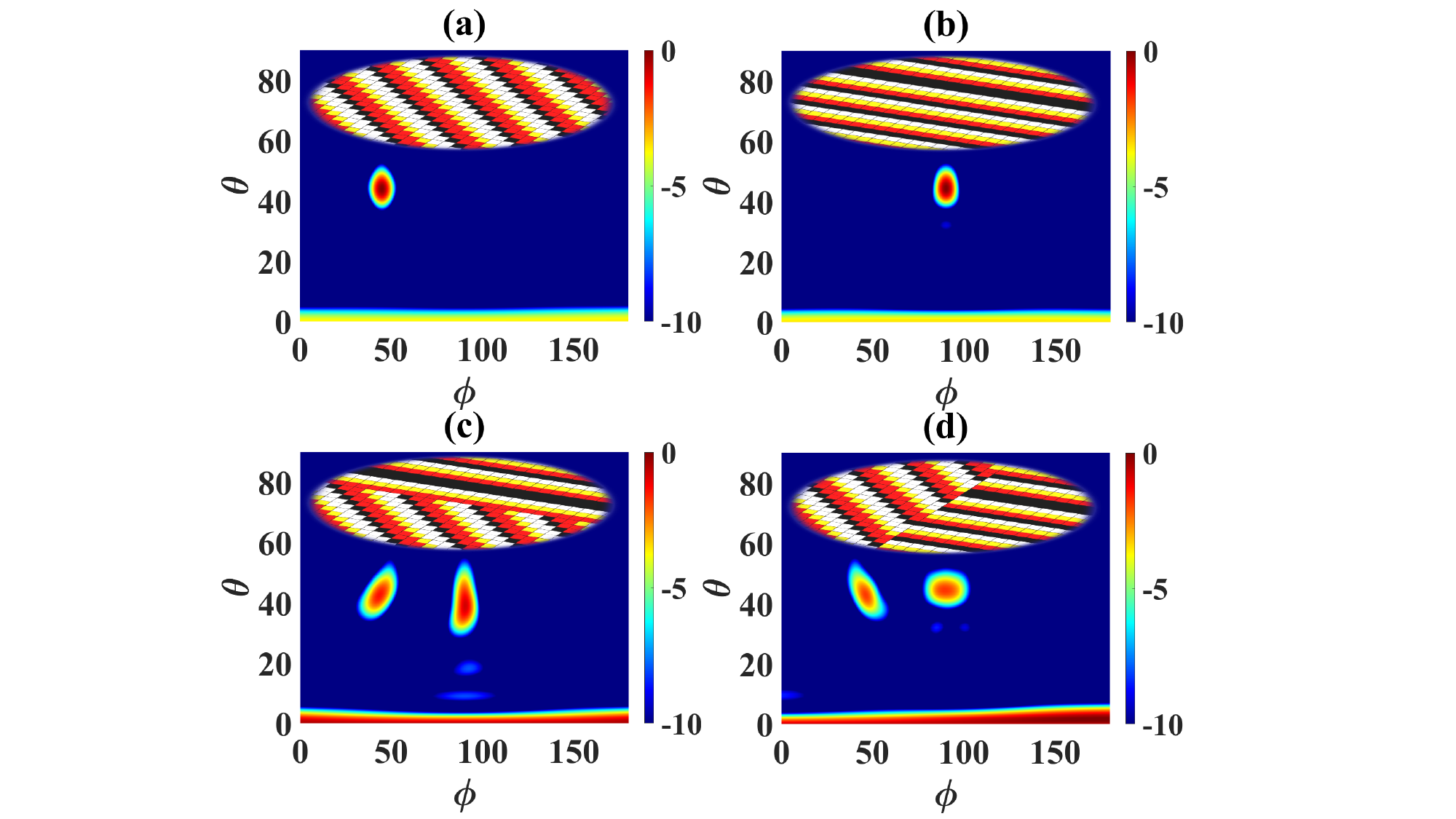}
\vspace{-.7cm}
\caption{Normalized E-field distribution in logarithmic scale (dB) with main beam radiation at (a) $\theta_r=\phi_r=\pi/4$ and (b) $\theta_r=\pi/4, \phi_r=\pi/2$ as single-user metasurfaces. Two main beam radiation, requires dividing the area of the \ac{MS} Row-wise (c) and Column-wise (d) to target both users.}
\label{fig:RC}
\end{figure}

A simple strategy to meet the 5G criteria is to communicate with all the receivers concurrently. So, instead of multiplexing in the time domain, we can partition the area of the \ac{MS} and assign per user segments. This segmentation process is equivalent to dividing the original \ac{MS} into tiles each may serve a single reflection \cite{9306896}, which inevitably follow by lowering the directivity. So, enlarging the \ac{MS} is essential to maintain the \ac{QoS} for multi-receiver scenarios. Figure \ref{fig:RC} illustrates an electromagnetic simulation performed with CST Microwave Studio \cite{CST}, in which the allocation of the \ac{MS} area amongst two beams reduces the directivity. 

\subsection{Amplitude and phase reconfiguration}
\label{sec:multi-theorem}
Independent amp/phs control of the unit cells has been proposed for multi-beam steering \cite{8753713}. However, amplitude reconfiguration not only applies losses in the reflection power but also requires sophisticated unit cell design and tuning mechanism to accurately control the amplitude and phase reflection simultaneously \cite{ashoor2020metasurface}. To radiate a pattern with multiple beams at several pairs of reflection angles (i.e., ($\theta_{r1},\phi_{r1}$), ($\theta_{r2},\phi_{r2}$)) single beam coding has to be modified. According to the desired direction of beams, one can calculate the relative phase profiles individually. Then, the principle of superposition allows to encapsulate the individual phase profiles \cite{Hashemi2016ABM,Wang2018,8753713}
\begin{equation}
\label{eq:addtwo}
Ae^{\Psi(\theta_{r},\phi_{r})}=e^{\Phi_1(\theta_{r1},\phi_{r1})}+e^{\Phi_2(\theta_{r2},\phi_{r2})}
\end{equation}
Even though the original profiles only have phases and the amplitude of these terms are unity, the out come of their summation has a matrix of amplitudes ($A$). In general, a summation form of arbitrary multi-beamforming results in
\begin{equation}
\label{eq:amp}
  \begin{aligned}[b]
\sum_{k=1}^{K}e^{j\Phi_{mn}(\theta_{rk},\phi_{rk})}=\Gamma_{mn} e^{j\Psi_{mn}}
  \end{aligned}
\end{equation}
where $\Phi_{mn}(\theta_{rk},\phi_{rk})$ is the phase gradient of $_{mn}$ unit cell for the kth-beam aiming $(\theta_{rk},\phi_{rk})$. The result of this summation is a term with both phase profile $\Psi_{mn}$ and amplitude profile $\Gamma_{mn}$. This means we can engineer a multi-beam radiation pattern by controlling the simultaneous amp/phs response of the unit cells.

\vspace{-.7cm}

\subsection{Optimization techniques}
While optimization methods can help us determine the best configuration for certain radiation patterns, they require extensive computing power and time. Since the number of possibilities of \ac{MS} configuration is huge, finding the optimized reconfiguration is not trivial. To exemplify the numbers, consider a grid of $10\times10$ unit cells. Next, as suggested in \cite{9109701}, we set $N_s=4$ states to code (i.e., setting specific phase and amplitude profiles) the \ac{MS}. Consequently, the overall configuration possibilities will be $4^{10\times10}$ and optimizing a large number of elements imposes a significant challenge. Nevertheless, instead of a full search over the entire design space, one can partition the unit cells into tiles, and optimize each tile in an offline design stage. In this case, only the element of the codebook is selected for online operations \cite{9306896}. Despite suggesting efficient algorithms based on e.g. convex optimization solvers to reduce the overhead of optimization and complexity of codebooks \cite{9350282}, the number of possibilities is still a fundamental issue. Therefore, we discard the investigation of this method. In Table \ref{com}, a qualitative comparison between different methods is made to give an overview on the advantages and disadvantages of each strategy.

\begin{table}[!htb]
    \centering
    \caption{Qualitative comparison between different strategies for multi-beam forming}
    \begin{tabular}{|p{1.35cm}|p{1.2cm}|p{1.2cm}|p{1.2cm}|p{1.2cm}|} 
    \hline
          \cellcolor{white} & \cellcolor{lightgray} \textbf{SDM} & \cellcolor{lightgray} \textbf{TDM} & \cellcolor{lightgray} \textbf{amp/phs} & \cellcolor{lightgray} \textbf{phase only} \\ \hline
          
         \cellcolor{lightgray} Requirement & \cellcolor{pink}large MS & \cellcolor{pink}fast reconfiguration & \cellcolor{pink} unit cell full control & \cellcolor{lime}low profile \\ \hline
         
         \cellcolor{lightgray} Side lobe level & \cellcolor{pink}high & \cellcolor{green}lowest & \cellcolor{lime}low & \cellcolor{yellow}moderate \\ \hline
         
         \cellcolor{lightgray} Directivity & \cellcolor{yellow}moderate & \cellcolor{green}highest & \cellcolor{yellow}moderate &  \cellcolor{lime}high \\ \hline
         
         \cellcolor{lightgray} latency & \cellcolor{lime}low & \cellcolor{yellow}moderate & \cellcolor{lime}low & \cellcolor{green}lowest \\ \hline
         
         \cellcolor{lightgray} Power &  \cellcolor{yellow}moderate & \cellcolor{pink}high &  \cellcolor{yellow}moderate & \cellcolor{lime}low \\ \hline

   \end{tabular}
    \label{com}
\end{table}

\section{Phase-Only Metasurface Coding for Anomalous Reflection in Multiple Directions}
\label{sec:Model}
In this section, we review the basics for single beam steering then we modify it for multi-beamforming. For anomalous reflection, \ac{MS} manipulates the reflected toward arbitrary direction. To this end, the reflection phase of each unit cell has to be controlled.
\subsection{Single beam/direction}
\label{sec:single}
Consider an impenetrable boundary interface under illumination of an incident plane wave. The incident wave vector $k_i$ with elevation angle $\theta_{i}$ and azimuth angle $\varphi_i$ can be written as
\begin{equation}\label{eq:incxy}
k_i = k_{ix} \hat{x} + k_{iy} \hat{y} + k_{iz} \hat{z}
\end{equation}
where {$k_{ix}, k_{iy}, k_{iz}$} are the wave vector coordinates
\begin{equation}\label{eq:incphi}
\begin{array}{l}
k_{ix}= k_i \sin\theta_{i} \cos\theta_{i} = k_0 n_0 \sin\theta_{i} \cos\theta_{i}\\
k_{iy} = k_i \sin\theta_{i} \sin\theta_{i} = k_0 n_0 \sin\theta_{i} \sin\theta_{i}\\
k_{iz} = k_i \cos\theta_{i} = k_0 n_0 \cos\theta_{i}
\end{array} 
\end{equation}
The same formulae can be derived for the reflected wave vector $k_r$ with the elevation angle $\theta_r$ and azimuth angle $\varphi_r$. 
\begin{equation}\label{eq:refphi}
\begin{array}{l}
k_{rx} = k_r n_r \sin\theta_{r} \cos\theta_{r} = k_0 n_0 \sin\theta_{r} \cos\theta_{r}\\
k_{ry} = k_r n_r \sin\theta_{r} \sin\theta_{r} = k_0 n_0 \sin\theta_{r} \sin\theta_{r}\\
k_{rz} = k_r n_r \cos\theta_{r} = k_0 n_0 \cos\theta_{r}
\end{array} 
\end{equation}
$k_0$ and $n_0$ are the wavenumber and refractive index of free space. In general, the direction of the reflected beam can be engineered by an appropriate linear phase profile \cite{PhysRevApplied.11.044024}. Assuming that the \ac{MS} imposes the phase profile $\Phi(x,y)$, we assign the virtual wave vector $\mathbf{k}_\Phi=\nabla\Phi=\partial_x \Phi\,\hat{x} + \partial_y \Phi\,\hat{y}$ ($\partial_x$ and $\partial_y$ denote partial derivatives). Applying the boundary conditions of the tangential components of the electromagnetic fields, the momentum conservation law for wave vectors can be expressed as \cite{8745693}
\begin{equation}\label{eq:mom}
\begin{array}{l}
k_{ix} + k_{\Phi x} = k_{rx}\\
k_{iy} + k_{\Phi y} = k_{ry}
\end{array} 
\end{equation}
from Equation \ref{eq:incphi}, \ref{eq:refphi} and \ref{eq:mom} we have
\begin{equation}\label{eq:dphi}
\begin{array}{l}
k_0 \sin{\theta_{i}}\cos{\theta_{i}} + \partial_x\Phi = k_0 \sin{\theta_{r}}\cos{\varphi_{r}}, \\
k_0 \sin{\theta_{i}}\sin{\varphi_{i}} + \partial_y\Phi = k_0 \sin{\theta_{r}}\sin{\varphi_{r}},
\end{array} 
\end{equation}
where $\partial_x\Phi$ and $\partial_y\Phi$ describe the imposed phase profiles in the $x$ and $y$ directions, respectively, and the subscripts $i$ and $r$ denote incident and reflected waves, respectively. Reflection position in the far-field is implied with pairs of angle variable $\theta_r$ and $\phi_r$ in a spherical coordinate system. Thus, the required phase profile for $mn$-th unit cell reads \cite{9109701}
\begin{equation}
\Phi_{mn}(\theta_r,\phi_r)=\frac{2\pi D_u (m\cos\varphi_r \sin\theta_r+n\sin\varphi_r \sin\theta_r)}{\lambda_0}
\label{eq:phimn}
\end{equation}
where $D_u$ is the length of a square unit cell, $\lambda_0$ is the wavelength in free space, and $m$ and $n$ are the indexes of the $mn$-th unit cell. According to the number of unit cell states $N_{s}$ and the phase gradient profile, we applied adaptive mapping such that the nearest available state registered to the unit cell. Using phase gradient described in Equation (\ref{eq:phimn}), we can encode the \ac{MS} to reflect the beam toward an arbitrary reflection angle $(\theta_r,\phi_r)$.

Assuming small unit cell size ($D_u<\lambda/2$), surface current distribution on each unit cell is approximately uniform. Based on Huygens principle, we can assume each unit cell is a point source. The total scattering field can be regarded as the superposition of the scattering wave from each unit cell \cite{Wang2014}
    \begin{equation}
\label{eq:E}
  \begin{aligned}[b]
E(\theta,\phi)= \sum_{n=1}^{N} \sum_{m=1}^{M} cos\theta e^{jk_0\zeta_{mn}}e^{j\Psi_{mn}}
  \end{aligned}
\end{equation}
where $\zeta_{mn}(\theta, \varphi)$ is the relative phase shift of the unit cells with respect to the radiation pattern coordinates, given by
\begin{equation}
\zeta_{mn}(\theta, \varphi) = D_u\sin{\theta}[(m-\tfrac{1}{2})\cos{\varphi}+(n-\tfrac{1}{2}) \sin{\varphi}]
\label{eq:phs}
\end{equation}
\subsection{Multiple beams/directions with energy conservation law}
\label{sec:multi-conversation}
Here, we propose a solution to discard the need for amplitude configuration. By considering the energy conservation law, in a closed system, the total energy from the impinging waves should be equal to the energy carried by the scattered beams. So, in the limit of lossless and fully reflective unit cell design only the phase of the reflected wave is shifted. Then, there must be an optimal reconfiguration profile with phase-only control by which we can engineer the desired multi-beamforming. Mathematically, by applying the right coefficients to the individual phase profile responsible for the singular beamforming, Equation (\ref{eq:addtwo}) become
\begin{equation}
\label{eq:addtwocons}
1e^{\Psi(\theta_{r},\phi_{r})}=a_1e^{\Phi_1(\theta_{r1},\phi_{r1})}+a_2e^{\Phi_2(\theta_{r2},\phi_{r2})}
\end{equation}
and equation (\ref{eq:amp}) become
\begin{equation}
\label{eq:ak}
\sum_{k=1}^{K}a_{mnk}e^{j\Phi_{mn}(\theta_{rk},\phi_{rk})}=1 e^{j\Psi_{mn}}
\end{equation}
This equation does not have a unique answer. Depending on the prioritization of the beams one can think of different configurations. Nevertheless, the simplest solution to satisfy Equation (\ref{eq:ak}) is to assume all of the beam amplitude are identical $a_{mnk}=a_{mn}$; $k\in[1,K]$. Then we can define this coefficient as the absolute value of the phase gradient summations.
\begin{equation}
\label{eq:Ak}
a_{mn}=\frac{1}{|\sum_{k=1}^{K}e^{j\Phi_{mn}(\theta_{rk},\phi_{rk})}|}
\end{equation}

from Equation (\eqref{eq:ak}) and (\eqref{eq:amp}), Equation (\eqref{eq:E}) reads
    \begin{equation}
\label{eq:Efinal}
  \begin{aligned}[b]
E(\theta,\phi)= \sum_{n=1}^{N} \sum_{m=1}^{M} cos\theta e^{jk_0\zeta_{mn}}a_{mn}\sum_{k=1}^{K}e^{j\Phi_{mn}(\theta_{rk},\phi_{rk})}
  \end{aligned}
\end{equation}
Now by using Equation (\eqref{eq:phimn}) and (\eqref{eq:Efinal}), we can implement a multi-beam radiation pattern. By selecting $D_u=\lambda/3$, we can ensure that the phase gradient is mapped on the \ac{MS} with acceptable resolution \cite{9109701}. In order to integrate the coding strategy, we used permittivity alternation to mimic the phase-shifting elements. We select a thickness of the dielectric slab as $l=\lambda_0/6$ where $\lambda_0=c/f$ and $f=28GHz$. Since this would be a homogeneous layout, the reflection phase from each unit cell can be analytically calculated as $e^{-2jkl}$. By imposing 4 different dielectric constants $\epsilon_r=[1,3.1,6.28,10.62]$, we coded the unit cells with 4 phase reflections as [$0$, $\pi/2$, $\pi$, $3\pi/2$] and unity amplitude.

\begin{figure}[!t]
\centering
\vspace{-0.4cm}
\includegraphics[trim={0 1.8cm 0 0cm},clip,width=1\columnwidth]{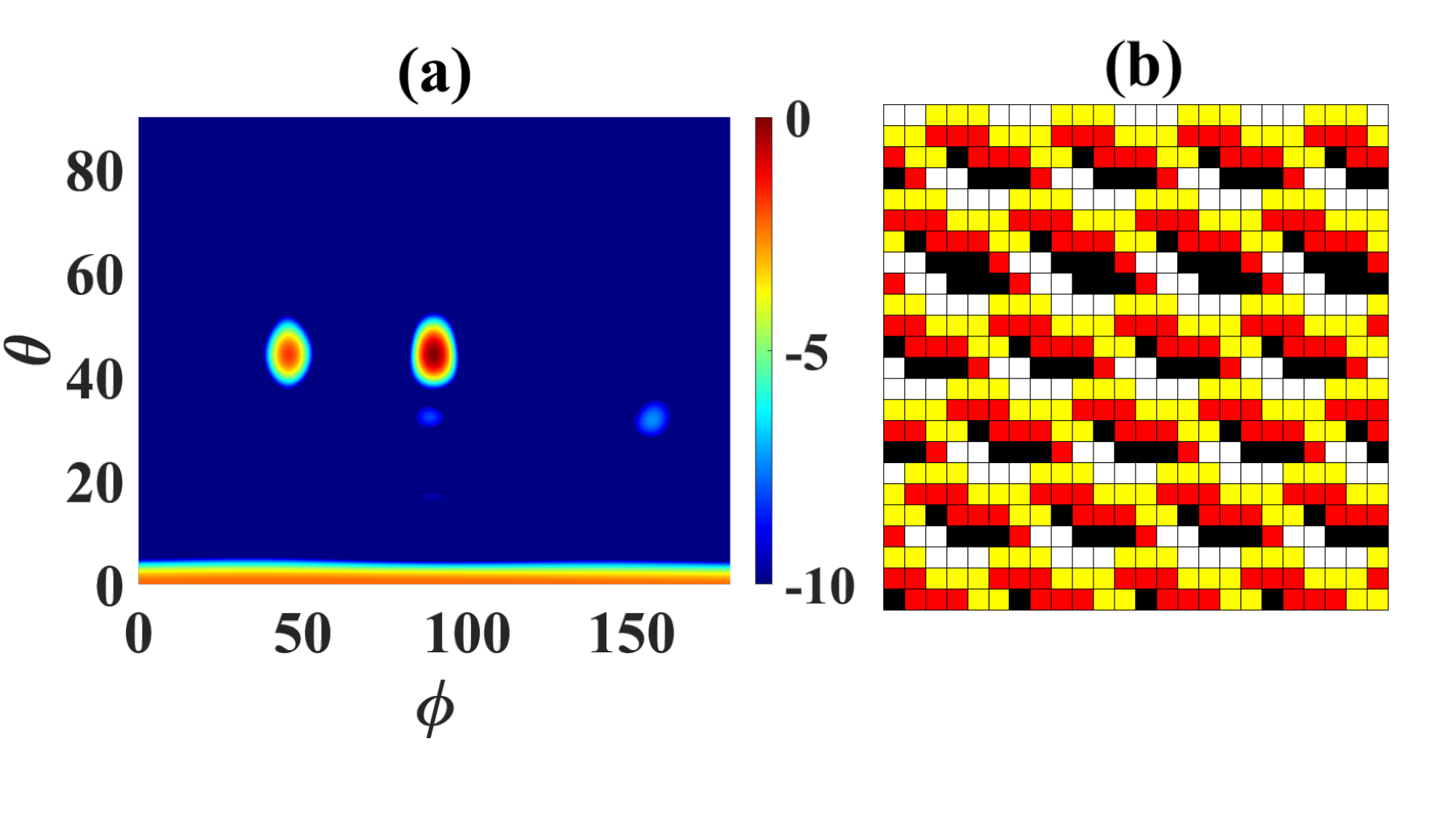}
\vspace{-0.7cm}
\caption{Normalized E-field distribution in logarithmic scale (dB) radiating with two beams at $\theta_r=\phi_r=\pi/4$ and $\theta_r=\pi/4, \phi_r=\pi/2$ (a) with respective phase gradient (b).}
\label{fig:OA2}
\end{figure}

Figure \ref{fig:OA2} (a,b), shows the radiation pattern and relative phase gradient of a square \ac{MS} with size of $D_m=8\lambda_0$. The wavelength is $\lambda_0\approx10 mm$ then MS size is $80\times80 mm$ and if the distance between the UEs and the MS is more than 1.5 meters, we can assume UEs are in the far-field zone. The obtained radiation pattern is improved compared to the spatial subdivision technique (Figure \ref{fig:RC} (c,d)).

\begin{figure}[!h]
\centering
\vspace{-0.4cm}
\includegraphics[trim={6cm 0cm 7cm 0cm},clip,width=1\columnwidth]{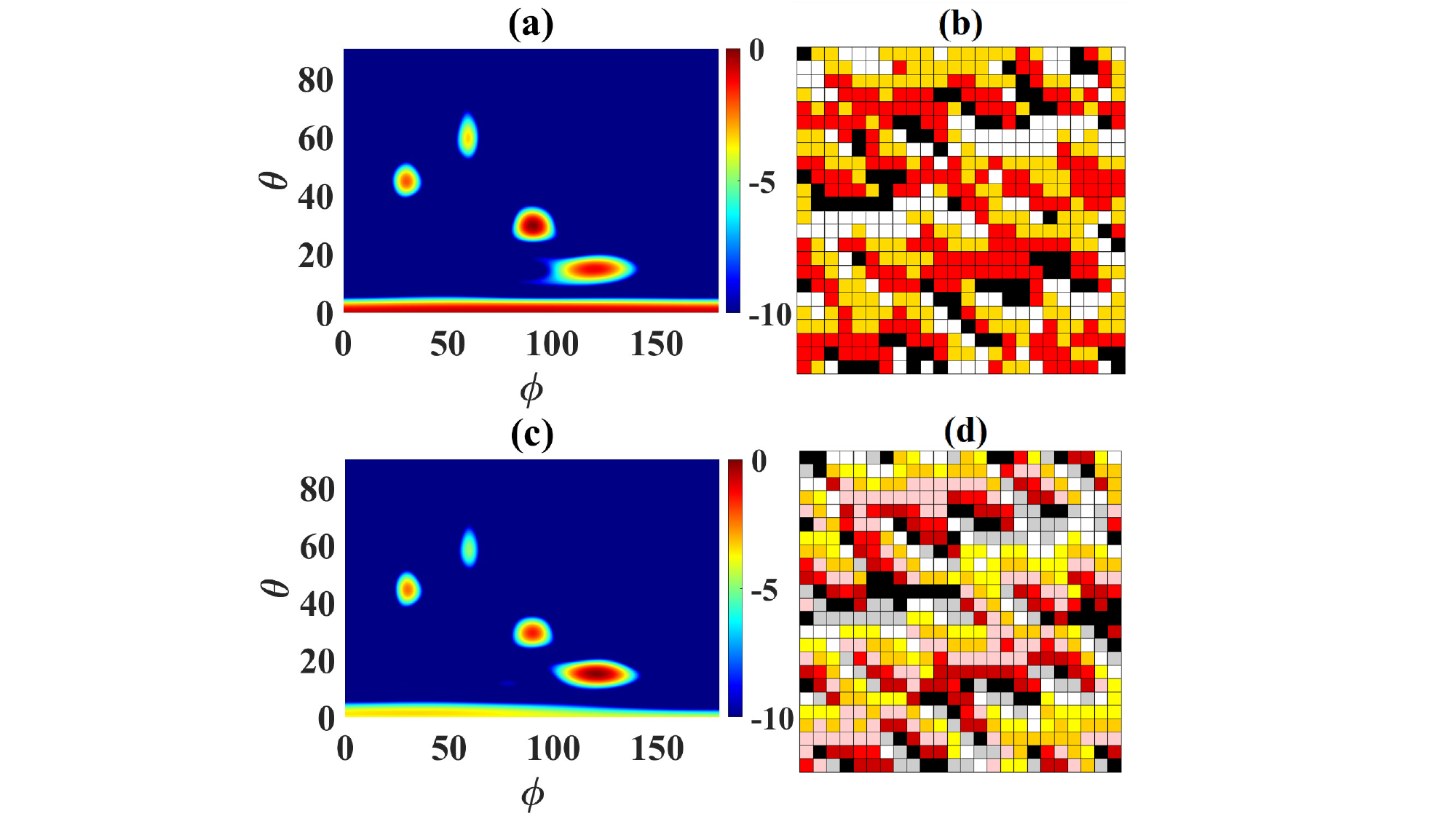}
\vspace{-0.7cm}
\caption{Normalized E-field distribution in logarithmic scale (dB) radiating with 4 beams at arbitrary positions with 4 and 8 states (a,c) and respective phase gradient with 4 and 8 number of states (b,d).} 
\label{fig:OA3}
\end{figure}

Since the dimension of the \ac{MS} is fixed, the generation of more beams decreases the directivity. The size of the \ac{MS} $D_m$, should be selected with respect to the number of beams. To provide complex radiation patterns with more beams, we need to impose the phase gradient with finer resolution. One way is using smaller unit cells which involves fabrication complexity and sophisticated configuration means. A proper strategy is to improve the mapping sequence by increasing the number of states ($N_s$). We checked the influence of $N_s$ in the case of 4 beams in Figure \ref{fig:OA3} such that (b) shows the phase gradient with 4 different colors representing $N_s=4$, (d) shows the phase gradient with 8 different colors representing $N_s=8$ and (a,c) are the respective radiation patterns. Apparently, in the bottom sub-figure (c) the Specular reflection at the normal direction ($\theta=0$) is 5 $dB$ weaker than the top sub-figure (a) which improves the efficiency of the system and decreases the back-scattering toward the transmitter.

\begin{figure}[!htb]
\centering
\vspace{-0.4cm}
\includegraphics[trim={1cm 6cm 1cm 6cm},clip,width=0.95\columnwidth]{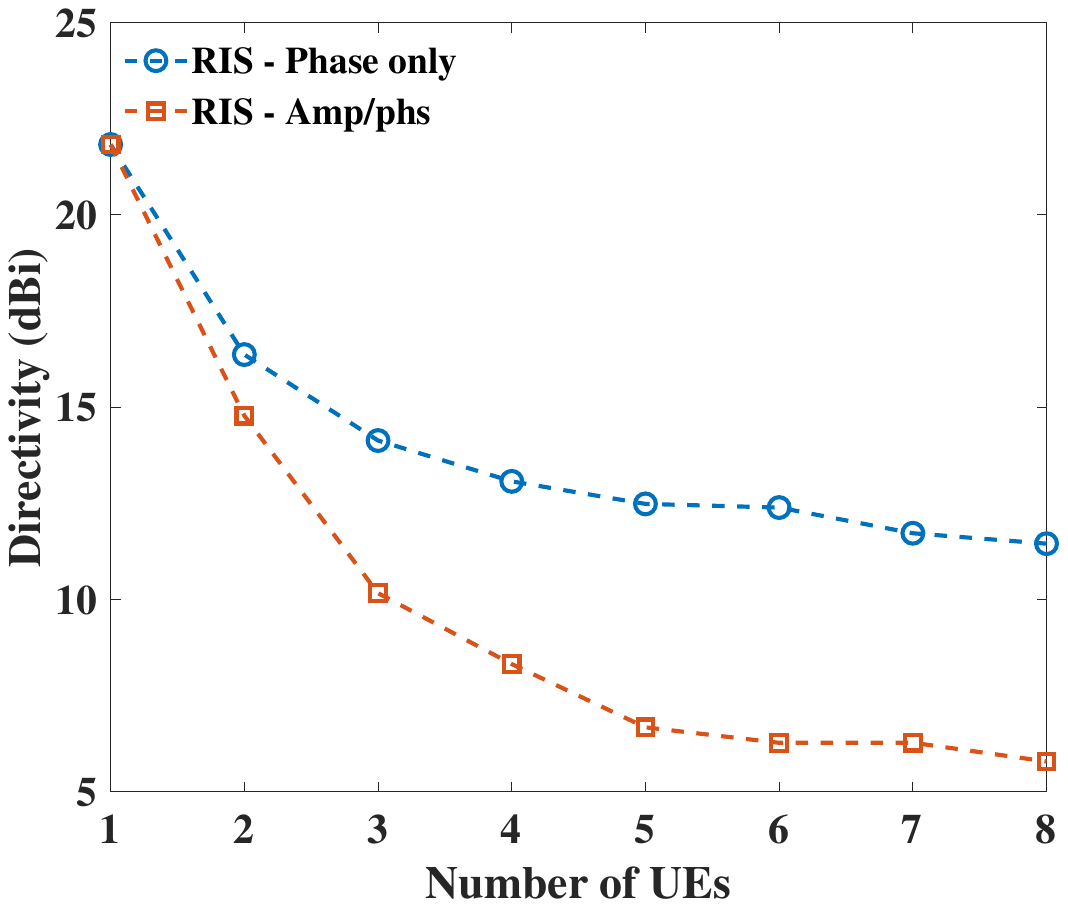}
\caption{Comparison of directivity vs number of users with phase only and amp/phs reconfiguration.}
\label{fig:Dir}
\end{figure}

Figure \ref{fig:Dir} compares the directivity of amp/phs and phase only reconfiguration with the same reference value of radiation intensity. Although amp/phase control yields very precise wavefronts, this does not mean that the obtained radiation power is high at the UE. The reason is amplitude configuration absorbs incident wave power at the unit cell level. So, the overall reflected power from the metasurface is weaker compared to phase only configuration. The difference of directivity between the two grows stronger as the number of UEs increases. In the following section, we build a system model and evaluate this difference in terms of channel capacity.

\section{Simulation setup}
\label{sec:scen}
To illustrate the efficacy of our \ac{MS} coding, we analyze its performance in the standard indoor and outdoor environments, as defined by \ac{3GPP} \cite{Report2018b}, and compare it with the current wireless network scenarios. The goal of this analysis is to compare the performance of the proposed \ac{MS} reconfiguration methodology as against the amp-phs reconfiguration method. Subsequently, in this section we first present the scenarios that will be evaluated, following which in Section \ref{sec:sysmod} we detail the system model utilized. Note that, while multiple research efforts \cite{9110915, 8796365,9541182,Ozdogan2020c, Bjornson2020b} do not consider realistic \ac{MS} operational characteristics such as directivity, we perform analysis by utilizing practical \ac{MS} performance parameters. These parameters have been determined using the technique described in Section \ref{sec:Model}. 

\subsection{Indoor Office Environment} \label{inh}

\begin{figure}[!htb]
\centering
\includegraphics[scale=0.5]{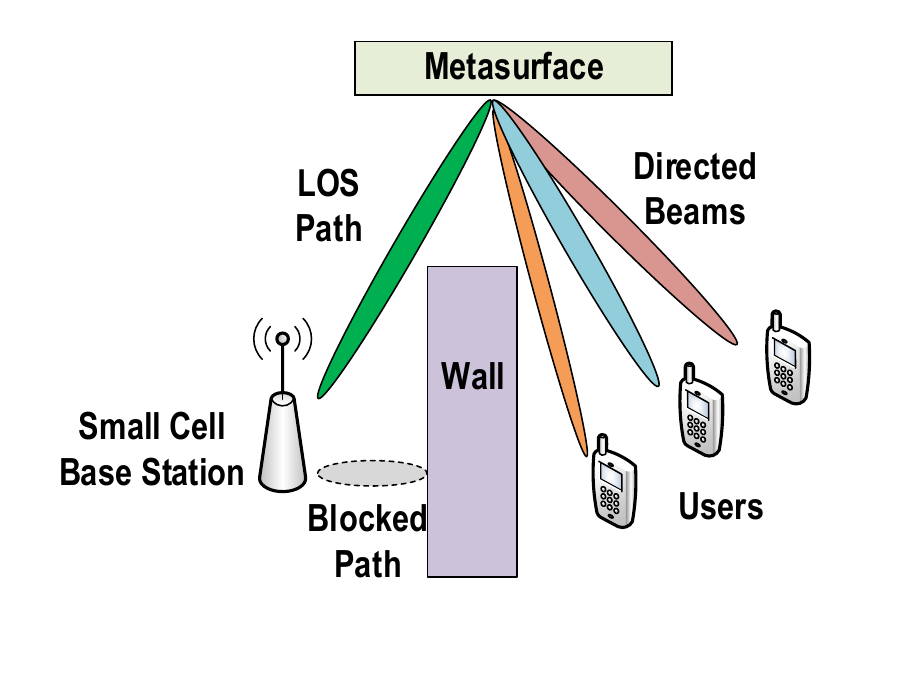}
\vspace{-0.4cm}
\caption{Indoor office environment propagation scenario. The small cell base station is blocked by an obstacle and the metasurface provisions a line of sight path for it towards all the users.}
\label{InH}
\end{figure}

The scenario corresponding to indoor office environments is presented in Fig. \ref{InH}. Characteristically, in such scenarios, the base stations (BSs) are low-powered transmitters, such as those for WiFi, etc, as compared to the cellular Access Points (APs). Moreover, the transmission path to the receivers can be blocked completely by obstacles (e.g., walls). Additionally, due to the density of obstacles, the propagation environment will be significantly impacted by multipath issues. The aforesaid impairments are further exacerbated for \ac{mmWave} frequencies \cite{jain2020user,jain2020mobility}. 

Hence, in Fig. \ref{InH}, the user equipment (UE) has the direct \ac{LoS} path from an small cell base station (SCBS) blocked by an obstacle. The SCBS-to-\ac{MS} link has a \ac{LoS} path. In addition, the \ac{MS}-to-UE link has a directed beam. We point out that, it is the \ac{MS} which provides a bridge (\ac{LoS} path) to the AP towards the UE, thus circumventing the complete blockage by the obstacle in between. 

\vspace{-0.4cm}

\subsection{Urban Micro Environment} \label{out}

\begin{figure}[!htb]
\centering
\includegraphics[scale=0.45]{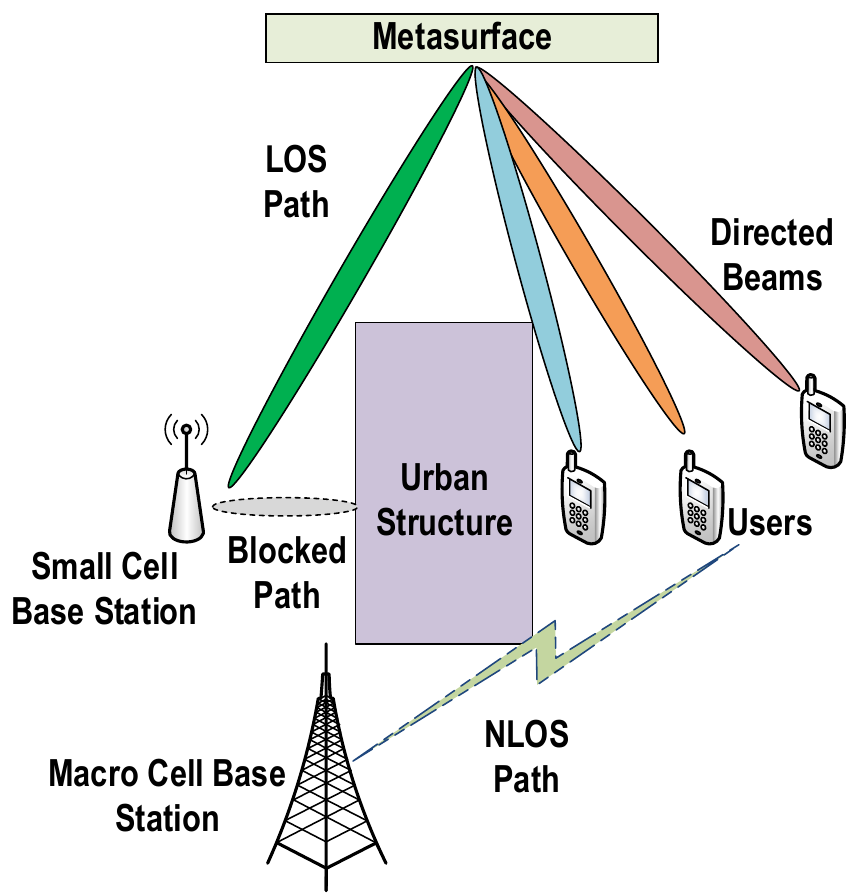}
\vspace{-0.4cm}
\caption{Urban macro environment propagation scenario, wherein the macro cell base satiation has a none line of sight path to the users due to an urban structure. However, the small cell base station is completely blocked by this urban structure. The metasurface provisions a line of sight path for the small cell base station to all the users.}
\label{Outdoor}
\end{figure}

The Urban Micro (UMi) environment, as shown in Fig. \ref{Outdoor}, consists of multiple BSs, i.e., the macrocells (MCBS) as well as the SCBSs, serving the users. In 5G and beyond scenarios, SCBSs are deployed to enhance the throughput, and hence, they will mostly operate upon the mmWave frequencies \cite{jain2020user,jain2020mobility}. On the other hand, MCBSs, or the anchor cells, will provide a more reliable connection to the users, thus maintaining coverage as well as supporting various dynamic scenarios \cite{MichaMaterniaNokia2016, Specification2018}. 

Consequently, while the SCBS will be blocked by the myriad obstacles present in a dense urban environment, such as that shown in Fig. \ref{Outdoor}, MCBSs will still have \ac{NLoS} path towards the users. 
Additionally, the SCBS has a LoS path through the \ac{MS} to the users, similar to the indoor environment in Section \ref{inh}.

\subsection{Broadcast}\label{buni}

As part of the analysis, in this work, for both the indoor and UMi scenarios, the broadcast mode of communications is evaluated. The broadcast (as well as multicast) mode enables the network to communicate the same information to multiple users at the same time. An important example of such an application is the video streaming service.

\subsection{System Model} 
\label{sec:sysmod}
Given the scenarios, we now discuss the system model for our evaluation. Firstly, we state that for both the indoor office environment and UMi scenarios, the BS and SCBS, respectively, communicate with the user through a LoS path facilitated by the \ac{MS}. Hence, the channel model for the aforesaid data path is represented as: 
\begin{align} \label{eq1}
    y_{sc} = \sqrt{G_{BM}G_{MU}}(\mathbf{g_{MU}^T\Theta h_{BM}})x_{sc} + \eta_{sc}
\end{align}
where $y_{sc}$ and $x_{sc}$ are the received and transmitted signals, respectively, and $\eta_{sc}$ denotes the additive white Gaussian noise with zero mean and variance (average noise power) $\sigma_{sc}^2$. Furthermore, $\mathbf{h_{BM}}$ and $\mathbf{g_{MU}}$ are the SCBS-to-\ac{MS} and \ac{MS}-to-UE channel coefficient vectors, respectively. Additionally, $\mathbf{\Theta}$ is the phase shift matrix that is formed by diagonalization of the vector of phase shifts applied at each element of the \ac{MS} on the received signal from the BS \footnote{In this paper, we primarily focus on the beamforming/beam-steering application towards multi-user environment, which is of significant importance for beyond 5G networks}. Finally, $G_{BM}$ is the transmit gain for the SCBS-to-MS path, i.e., SCBS transmit gain, and $G_{MU}$ is the transmit gain for the MS-to-UE path, i.e, MS transmit gain. This can be obtained by computing the product of the directivity values in figure \ref{fig:Dir} and the antenna efficiency (c.f. Table \ref{params}).

Next, the channel model for the MCBS to UE path in the UMi scenario (see Section \ref{out}) is defined as: 
\begin{align} \label{eq2}
    y_{mc} = \sqrt{G_{mc}}h_{mc}x_{mc} + \eta_{mc}
\end{align}
where $y_{mc}$ and $x_{mc}$ are the received signal at UE from MCBS and transmitted signal from MCBS to UE, respectively. The channel coefficients for the MCBS to UE channel is represented by $h_{mc}$, with the additive white Gaussian noise represented as $\eta_{mc}$ which has zero mean and variance (average noise power) $\sigma_{mc}^2$. Further, $G_{mc}$ is the MCBS transmit gain. From Eqs. (\ref{eq1}) and (\ref{eq2}), the overall received SNR at UE for the indoor scenario is determined as: 
\begin{align} \label{InhSNR}
    \Delta_{InH} = \frac{\left|\left|y_{sc} \right|\right|^{2}}{\sigma_{sc}^{2}}
\end{align}
whereas the received SNR at the UE for the UMi scenario is expressed as: 
\begin{equation} \label{OutSNR}
    \Delta_{UMi} =  
    \begin{cases} 
      \frac{\left|\left|y_{sc} \right|\right|^{2}}{\sigma_{sc}^{2}} &  \text{SNR at UE from SCBS},  \\
      \frac{\left|\left|y_{mc} \right|\right|^{2}}{\sigma_{mc}^{2}} & \text{SNR at UE from MCBS} \\
    \end{cases}
\end{equation}

Next, the maximum achievable throughput for the users in both the indoor and UMi scenarios can be defined by the Shannon-Hartley theorem as follows: 
\begin{align} \label{maxTP}
    R = B\log_{2}(1+SNR) 
\end{align}
where, $R$ is the maximum achievable throughput for a user, $B$ is the allocated bandwidth by a base station (BS/SCBS/MCBS), and $SNR$ is the Signal-to-noise ratio at the receiver from a given base station. Hence, from Eqs. (\ref{InhSNR}), (\ref{OutSNR}) and (\ref{maxTP}), the maximum throughput for a user in the Indoor environment, i.e., $R_{inh}$, is given as: 
\begin{align} \label{InhTP}
    R_{inh} = B_{inh}\log_{2}(1+\frac{\left|\left|y_{sc} \right|\right|^{2}}{\sigma_{sc}^{2}}) 
\end{align}
where $B_{inh}$ is the bandwidth allocated to the user by the BS. On the other hand, the maximum achievable throughput for the UMi scenario, as shown in Fig. \ref{Outdoor}, is:
\begin{align} \label{OutTP}
    R_{UMi} = B_{sc}\log_{2}(1+\frac{\left|\left|y_{sc} \right|\right|^{2}}{\sigma_{sc}^{2}}) + B_{mc}\log_{2}(1+\frac{\left|\left|y_{mc} \right|\right|^{2}}{\sigma_{mc}^{2}})
\end{align}
where, $R_{UMi}$ is the achievable throughput, and $B_{sc}$ and $B_{mc}$ are the allotted bandwidths to the user from the SCBS and MCBS, respectively. 

However, to compute the received signal powers, i.e., $\left|\left|y_{sc} \right|\right|^{2}$ and $\left|\left|y_{mc} \right|\right|^{2}$ in Eqs. (\ref{InhTP}) and (\ref{OutTP}), we evaluate the pathloss from the SCBS and MCBS to the UE. Subsequently, we utilize the the link budget formula in Equation (\eqref{linkb}) to compute the received signal power as follows,   
\begin{align}\label{linkb}
    P_r = P_t + G_t + G_r - PL - L_o
\end{align}
where $P_r$ is the received power, $P_t$ is the transmitted power, $G_r$ is the gain at the receiver antenna, $G_t$ is the gain at the transmit antenna ($G_{BM}$, $G_{MU}$ and $G_{mc}$), $PL$ is the scenario dependent path loss and $L_o$ are the other losses incurred at the transmitter and receiver feed, and other mismatches, etc. Note that, in this work we ignore the other losses $L_o$ for the sake of simplicity. In addition, we define the pathloss models \footnote{In this work, we assume that there is no shadow fading. Hence, in eqs. (17)-(20) we do not introduce the shadow fading parameter.}, based on the CI and 3GPP models \cite{jain2020user, 8386686, Report2018b,MichaMaterniaNokia2016}, as follows: 
\begin{align} \label{plumi}
    PL_{UMi} = 20\log_{10}(\frac{4\pi f}{c}) + 10n\log_{10}(d_{3D}) 
\end{align}
\begin{equation}\label{plLOS}
    PL_{inh-LOS} = 32.4 + 20\log_{10}(f) + 17.3\log_{10}(d_{3D}) 
\end{equation}
\begin{equation}\label{plNLOS1}
    PL_{inh-NLoS} = max(PL_{inh-LoS},PL'_{inh-NLoS})
\end{equation}
\begin{equation}\label{plNLOS2}
    PL'_{inh-NLoS} = 38.3\log_{10}(d_{3D}) + 17.30 + 24.9\log_{10}(f) 
\end{equation}
where, $PL_{UMi}$, $PL_{inh-LoS}$, $PL_{inh-NLoS}$, and $PL'_{inh-NLoS}$ are the pathloss for the UMi scenario (for both LoS and NLoS setups), indoor office LoS scenario and the NLoS scenarios, respectively. In addition, $c$ is the speed of light, $f$ is the central frequency of operation and $d_{3D}$ is the 3D distance between the transmitter and receiver. 

Next, for the received signal power computation in Equation (\eqref{linkb}), the transmit power, transmitter gain, and receiver gain are required. While these parameters for the SCBS, MCBS and UEs are readily available through existing literature \cite{MichaMaterniaNokia2016, jain2020user}, a practical and realistic estimate of transmitter gain for an \ac{MS} in the presence of single and multiple receivers is largely missing from the current literature. Note that in our study we assume the receiver gain of the \ac{MS} as 0 dBi. Moreover, for the SCBS to UE pathloss value in eq. (16), a two step process has to be followed, i.e., a pathloss from SCBS to MS and another one from MS to UE has to be calculated. Thus, for the SCBS to MS pathloss computation the values from \cite{MichaMaterniaNokia2016, jain2020user} and the 0 dBi MS receive gain assumptions are utilized. Next, for the MS to UE pathloss calculation, the transmit power is the actual received power from the SCBS to MS path and the MS transmit gain is $G_{MU}$ which we obtain from directivity values in Figure \ref{fig:Dir}, as mentioned earlier. The other parameters are again used based on \cite{MichaMaterniaNokia2016, jain2020user}.

\begin{table*}[!htb]
    \centering
    \vspace{-0.4cm}
    \renewcommand{\arraystretch}{1.1}
    \caption{System model parameters}
    \begin{tabular}{|>{\centering\arraybackslash}m{3cm}|>{\centering\arraybackslash}m{2cm}|>{\centering\arraybackslash}m{2cm}||>{\centering\arraybackslash}m{3cm}|>{\centering\arraybackslash}m{2cm}|>{\centering\arraybackslash}m{2cm}|}
    \hline
         \textbf{Parameter} & \textbf{Indoor Office Environment} & \textbf{UMi Environment}&\textbf{Parameter} & \textbf{Indoor Office Environment} & \textbf{UMi Environment}  \\ \hline
         SCBS operating frequency & 28 GHz & 28 GHz & MCBS operating frequency & -- & 3.55 GHz \\  \hline
         Transmit Power SCBS & 37 dBm & 37 dBm & Transmit Power MCBS & -- & 49 dBm \\ \hline
         Transmit Gain SCBS & 30 dBi & 30 dBi & Transmit Gain MCBS & -- & 17 dBi \\ \hline 
         \ac{MS} Receive Gain & \multicolumn{2}{c||}{0 dBi} & \ac{MS} directivity & \multicolumn{2}{c|}{cf. Fig. \ref{fig:Dir} } \\ \hline
         \ac{MS} efficiency & \multicolumn{2}{c||}{0.9} & Noise spectral density & \multicolumn{2}{c|}{-174 dBm/Hz} \\ \hline 
         Pathloss exponent for SCBS NLoS & 3.8 & 3.2 & Pathloss exponent for SCBS LoS & 1.7 & 2.1 \\ \hline
          Pathloss exponent for MCBS NLoS & -- & 2.9 & Pathloss exponent for MCBS LoS & -- & 2.0 \\ \hline
    \end{tabular}
    \label{params}
\end{table*}

Lastly, we introduce Table \ref{params}, wherein we detail the other system model parameters/settings for the indoor and UMi scenarios. 
Specifically, we list the scenario parameters such as BS/SCBS/MCBS heights, pathloss exponents, shadow fading standard deviation, transmit power, and gains according to 3GPP specifications \cite{Report2018b} and recent research works such as \cite{8386686,jain2020user}.

\section{Discussion} 
\label{sec:eval}

We evaluate the performance of our \ac{MS} driven network in both indoor offices (Fig. \ref{InH}) and UMi scenarios (Fig. \ref{Outdoor}) based on the system models and parameters defined in Section \ref{sec:sysmod}. We perform the evaluation based on the channel capacity analysis and provide corresponding insights.       

\subsection{Indoor office scenario} \label{inheval}
\begin{figure}[!htb]
\centering
\vspace{-0.5cm}
\includegraphics[trim=220 0 220 0, clip, width = \columnwidth]{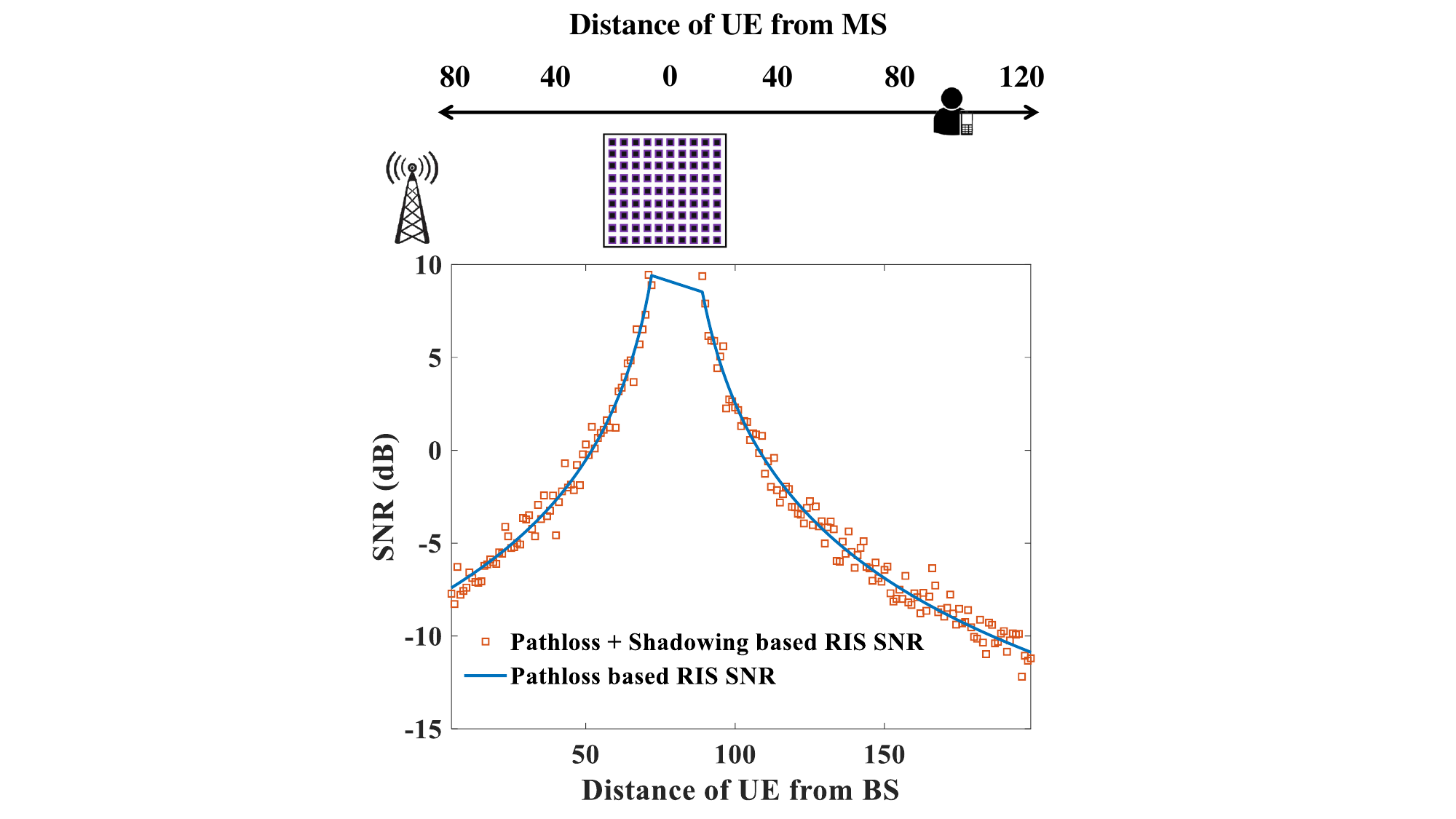}
\caption{Signal to noise ratio vs distance of the user from the base station (the RIS at 80 meters distance from base station). As shown while the base station and the RIS are fixed whereas the user is moving away from the base station up to 200 meters.}
\label{SNRDist80}
\end{figure}

Before delving deeper into the channel capacity analysis, it is imperative to understand the behavior of the \ac{MS}s in the wireless environment. By behavior, we mean that the SNR profile of the wireless channel corresponding to the reflected path from the \ac{MS}. This is an essential step, as it highlights the channel properties of the reflected path from the \ac{MS}. Thus, through Fig. \ref{SNRDist80} we present an analysis of the SNR and pathloss characteristics of the wireless channel (reflected path from \ac{MS}) in an indoor office environment. In the simulation setup, the \ac{MS} was placed at distance of 80 meters from the BS. In addition, a single user was considered and moved from 1m up to 200m from the BS. The BS is located at $(x,y)=(10,0)$ with 25 meters height and the MS is at $(x,y)=(10,80)$ with 5 meters height and UE is moving in 1D direction along the y axis with the same height and longitude ($x$). Subsequently, the received SNR was computed for each of the location combinations of the user and \ac{MS}. The apparent jump in the SNR diagram in Fig. \ref{SNRDist80} is there to skip the near-field zone where our assumptions do not hold.

From the SNR profile, it is evident that the received SNR is highest when the UE is close to the \ac{MS}. Hence, the overall SNR degradation scales up accordingly as the distance of the user increases from the \ac{MS}. So, the SNR gets weaker when UE is moving toward the BS because most of the signal reaching the receiver is a result of the action of the MS. This gives an initial assessment of the fact that \ac{MS}s are more effective for close-range communications. Note that, while we have analyzed this for the indoor environment, the observed pathloss phenomenon is also valid for the outdoor scenario. Following this observation, we now present the channel capacity analysis for the indoor environment. 

\begin{figure}[!htb]
\centering
\includegraphics[trim={1cm 6cm 1cm 6cm},clip,width=0.95\columnwidth]{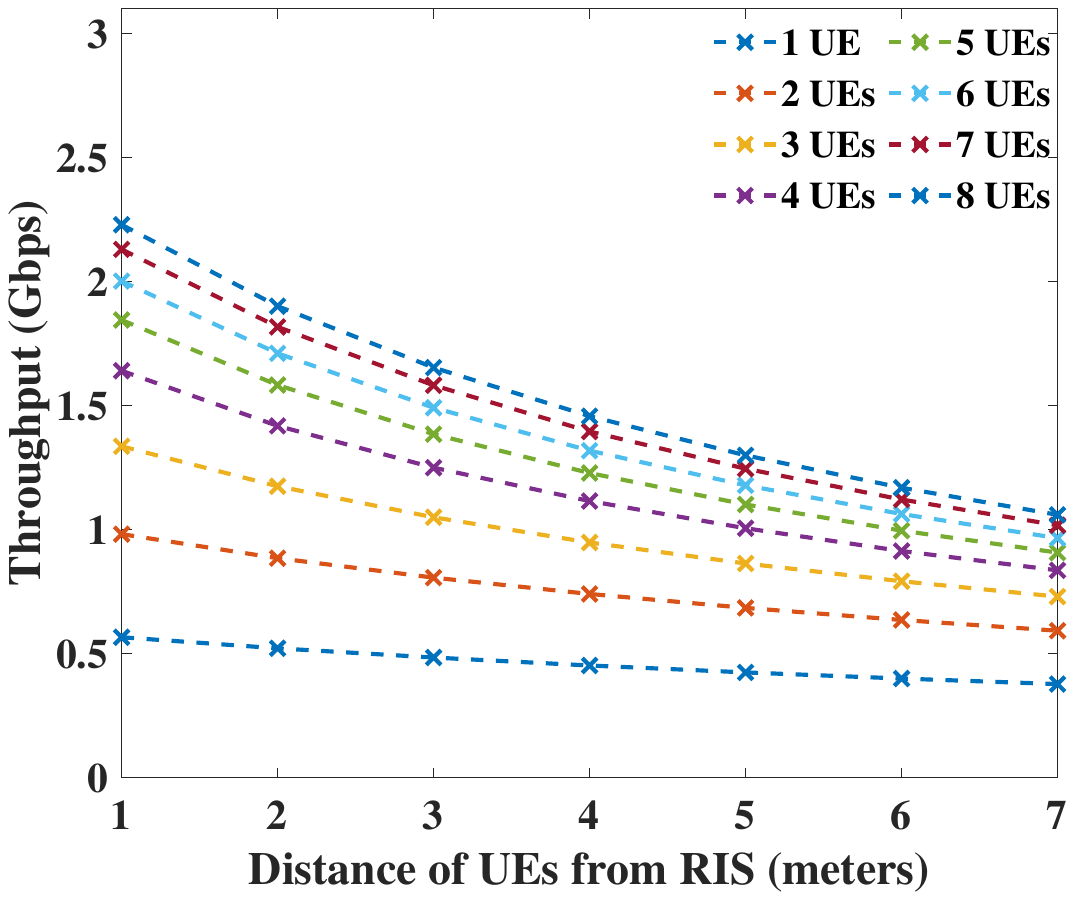}
\includegraphics[trim={1cm 6cm 1cm 6cm},clip,width=0.95\columnwidth]{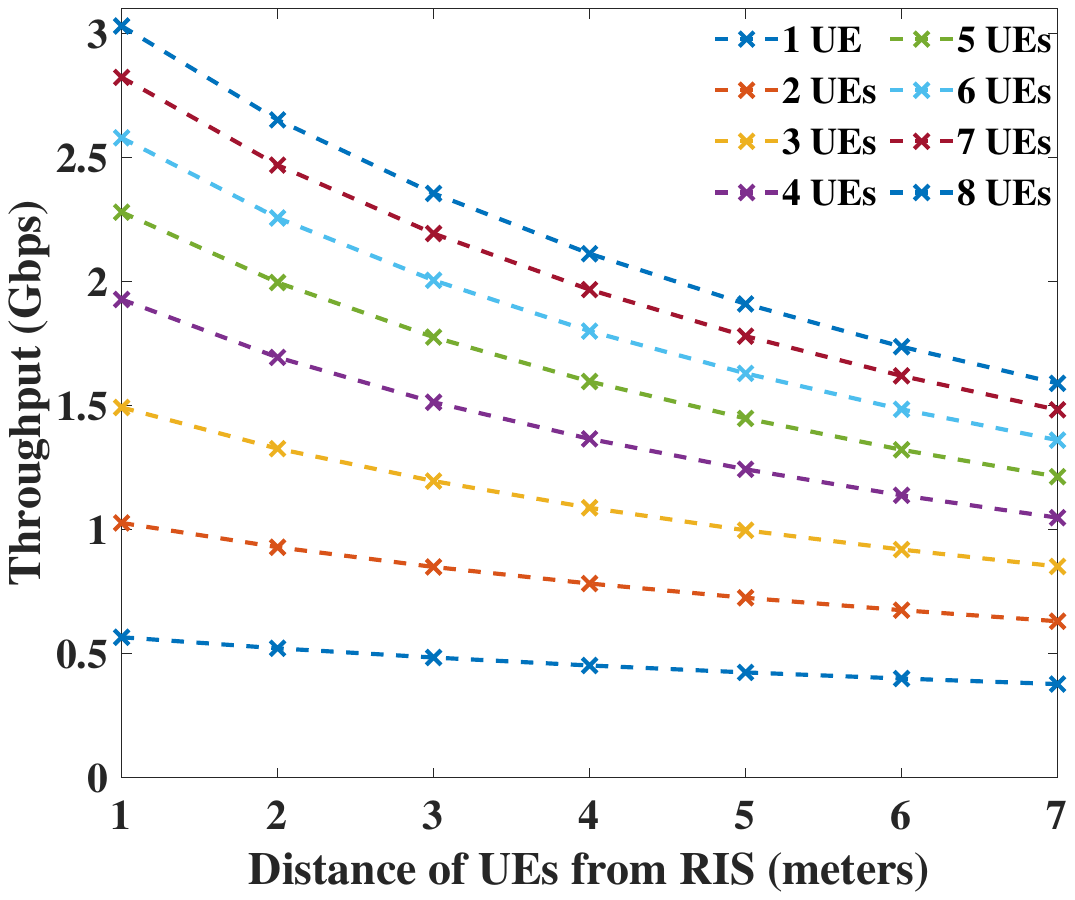}
\caption{Comparison of the throughput for amp/phs (top) and phase only reconfiguration (bottom) in indoor scenario versus distance to RIS.}
\label{inhCapChan}
\end{figure}

The SCBS uses an antenna with $30$ dBi directivity that operates at $28$ GHz is located at $(x,y)=(0,0)$ with 10 meters height. The RIS is at $(x,y)=(10,100)$ with 5 meters height and UEs are respectively $[5,4,3,3,4,5,6,7]$ meters away from the RIS with same height. The RIS also operates at $28$ GHz, its directivity is reported in Fig. \ref{fig:Dir} with phase only configuration and amp/phs configuration. Figure \ref{inhCapChan} compares the throughput from each reconfiguration method versus the distance of UEs from the RIS. Note that this distance is an additional parameter to the UEs location ($[5,4,3,3,4,5,6,7]$) and the UEs are for sure in the far-field zone.

Due to the propagation loss, throughput degrades when the UEs are further away from the RIS. As the number of UEs increases, the throughput grows but if we divide it between the UEs, in fact throughput for each user drops down. For instance, this system provides 2 Gbps throughput for 6 UEs thus each user has 0.33 Gbps throughput. The proposed phase only configuration provides more throughput compare to amp/phs configuration. The difference increases by the number of UEs such that for 8 UEs the throughput increases by around 0.8 Gbps.

\subsection{UMi scenario} \label{umieval}
We now present the channel capacity analysis for the UMi environment given the broadcast scenario. MCBS uses an antenna with $17$ dBi directivity that operates at $3.55$ GHz is located at the center of the coordinate system $(x,y)=(0,0)$. The altitude of the antenna is 25 meters. The SCBS is located at $(x,y)=(10,2000)$ with 10 meters height. The RIS is at $(x,y)=(20,2100)$ with 5 meters height and UEs are respectively $[5,4,3,3,4,5,6,7]$ meters away from the RIS with same height. SCBS and RIS characteristics are same as before.

\begin{figure}[!htb]
\centering
\includegraphics[trim={1cm 6cm 1cm 6cm},clip,width =0.95\columnwidth]{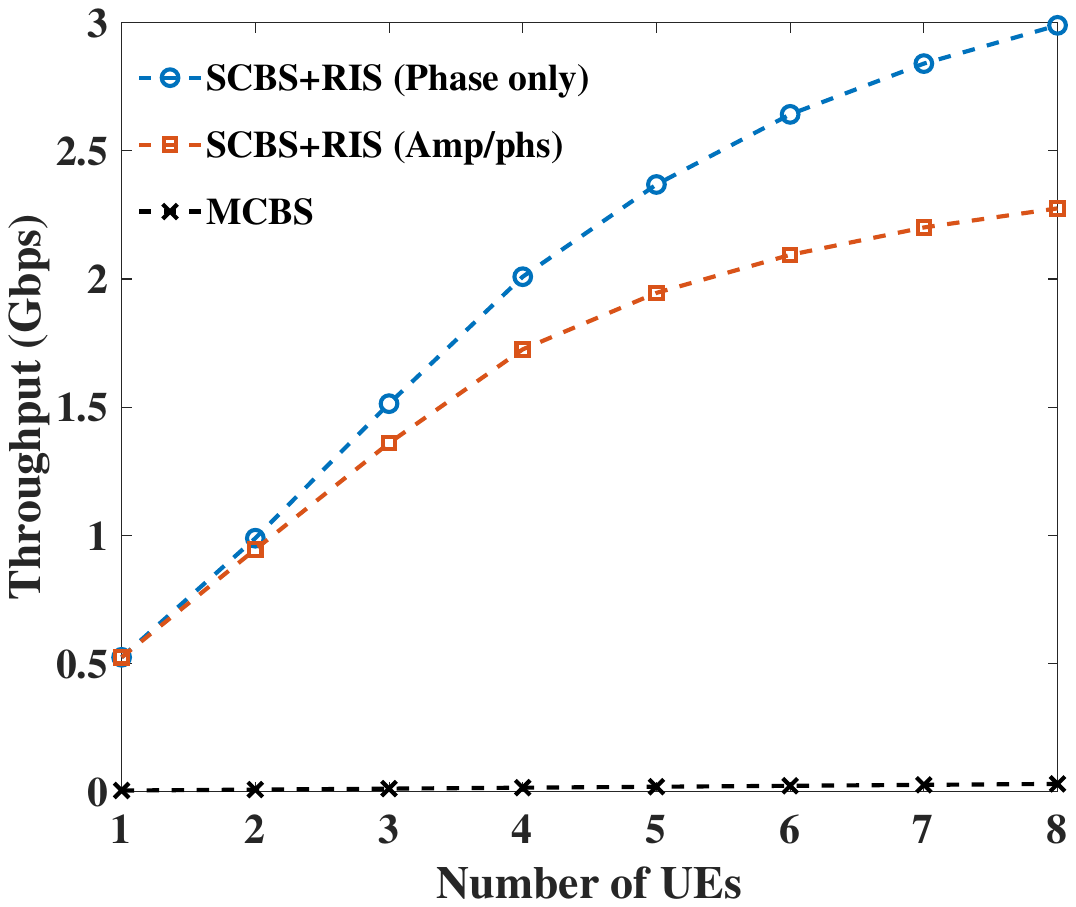}
\vspace{-0.2cm}
\caption{Urban micro environment channel capacity analysis.}
\label{UmiCapChan}
\end{figure}

\begin{figure}[!htb]
\centering
\includegraphics[trim={1cm 6cm 1cm 6cm},clip,width =0.95\columnwidth]{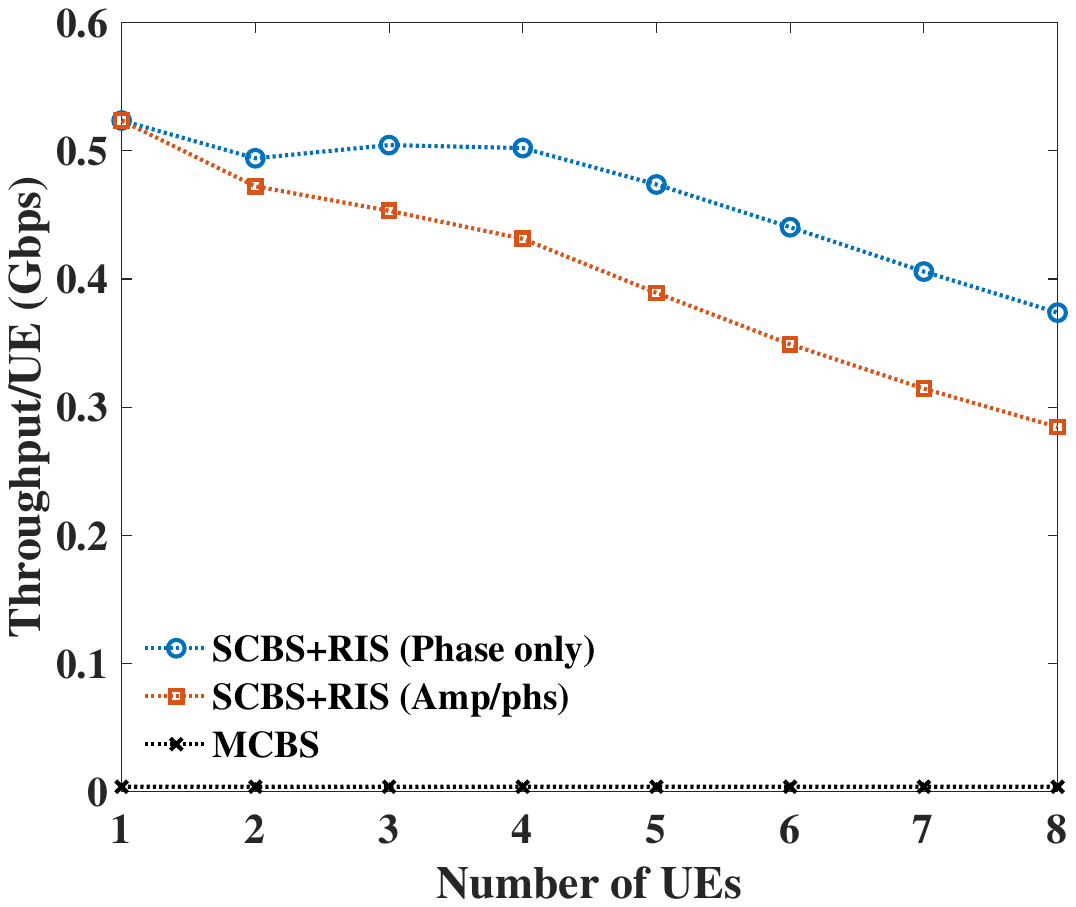}
\vspace{-0.2cm}
\caption{Urban micro environment channel capacity analysis per UE.}
\label{UmiCapChanUE}
\end{figure}

Figure \ref{UmiCapChan} compares the throughput of the system with and without RIS. Direct throughput from the MCBS is negligible compared to the throughputs achieved when utilizing the RIS (MS). This is because of the NLOS path that the MCBS faces in a UMi scenario, as well as the low bandwidth that it can provision to the users. However, the SCBS empowered with RIS provides overcomes the blockage and provisions nearly 0.5 Gbps for one user. The overall throughput (channel capacity) gets higher as the number of UEs increases. On the other hand, according to Fig. \ref{UmiCapChanUE}, the throughput per UE gets lower by increasing the number of UEs. Similar to in the indoor scenario, the phase only configuration provides higher input compared to amp/phs configuration for both channel capacity and throughput per UE.


\section{Conclusion}
\label{sec:conclusions}
The proposed method avails optimum theory for space-multiplexing to reconfigure the \ac{MS}. Concretely, this proposal provisions independent control over the radiation pattern lobes by which multi-user communication links can be established. Subsequently, the analysis shows that the \ac{MS} based system provides the best performance when the \ac{MS} is located close to the users. Further, we observed promising performance for indoor office and UMi environments given the broadcast mode of operation. Specifically, in the indoor office scenario, we observe that if the users are within 1-2m of the \ac{MS}, then at least 0.5 Gbps of data rate can be experienced by the users (with a peak data rate of $\sim$ 2.2 Gbps). Next, for the UMi scenario, we observed that the \ac{MS} based system provisions more than one order of magnitude more channel capacity in the presence of 7 users compared to MCBS communication. Hence, through this work, we have shown the efficacy and effectiveness of the designed \ac{MS} for 5G and beyond scenarios with multi-user applications.

\section*{Acknowledgments}
This work has been supported by the European Commission in part by the H2020 RISE-6G Project under Grant 101017011 and in part by the EPSRC under Grant EP/V048937/1. The work of Gabriele Gradoni was supported by the Royal Society under Grant INF$\backslash$R2$\backslash$192066.

\vspace{-1.3cm}

\begin{IEEEbiography}[{\includegraphics[width=1in,height=1.25in,clip,keepaspectratio]{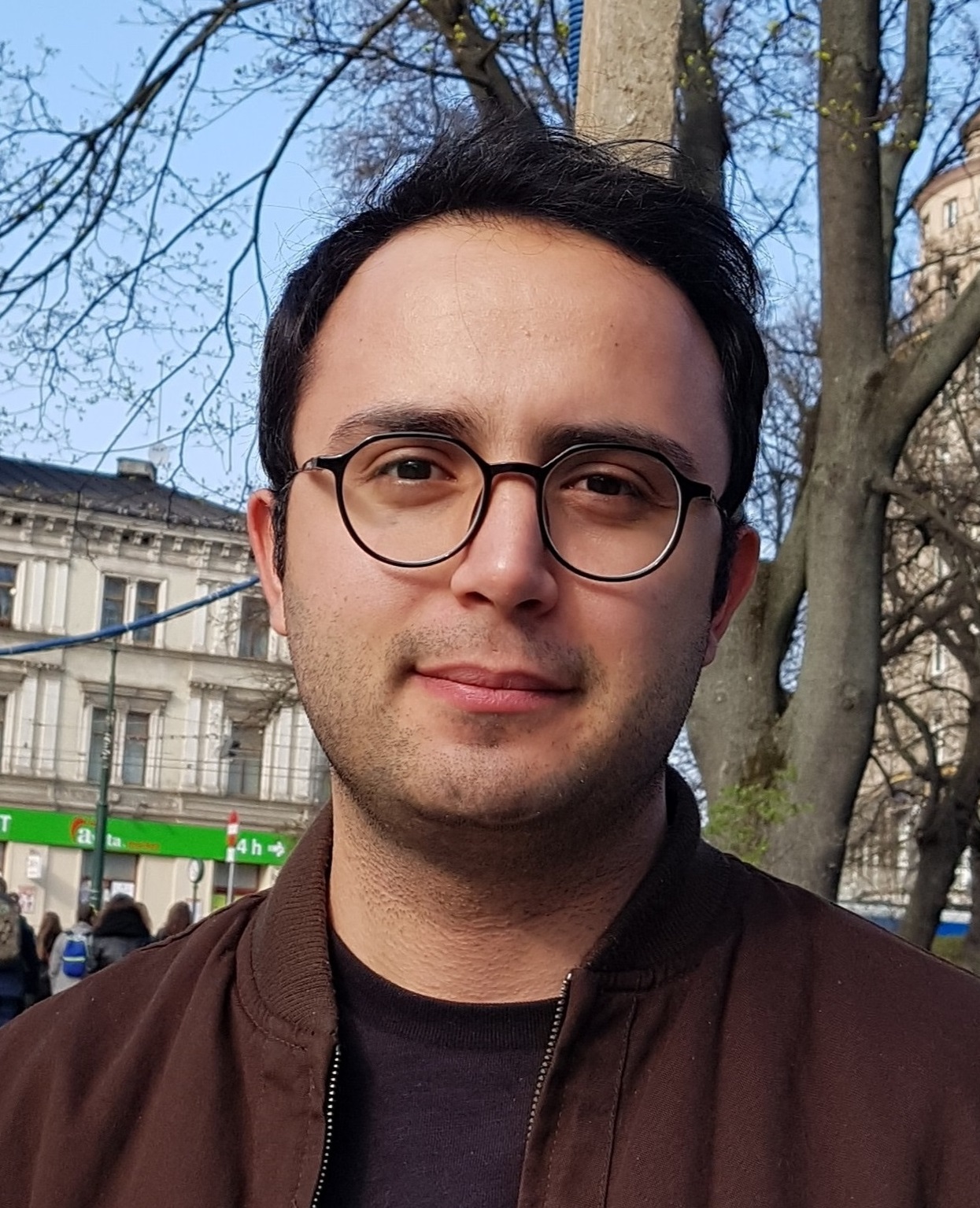}}]{Hamidreza Taghvaee} received the M.Sc. in Telecommunication Engineering majored in Field$\&$Wave from K. N. Toosi University of Technology, Iran, in 2016. He received the cum laude and the Ph.D. degree in Computer Science Engineering from the Universitat Politècnica de Catalunya, Spain, in 2021. He was a Visiting Researcher with the Department of Electronics and Nanoengineering, Aalto University, Finland, in 2020. Currently, he is a Research Fellow with the George Green Institute for Electromagnetics Research, University of Nottingham, United Kingdom. His main research interests include Electromagnetics, Metamaterials, Wireless Communications and Antennas.
\end{IEEEbiography}
\vspace{-1cm}

\begin{IEEEbiography}[{\includegraphics[width=1in,height=1.25in,clip,keepaspectratio]{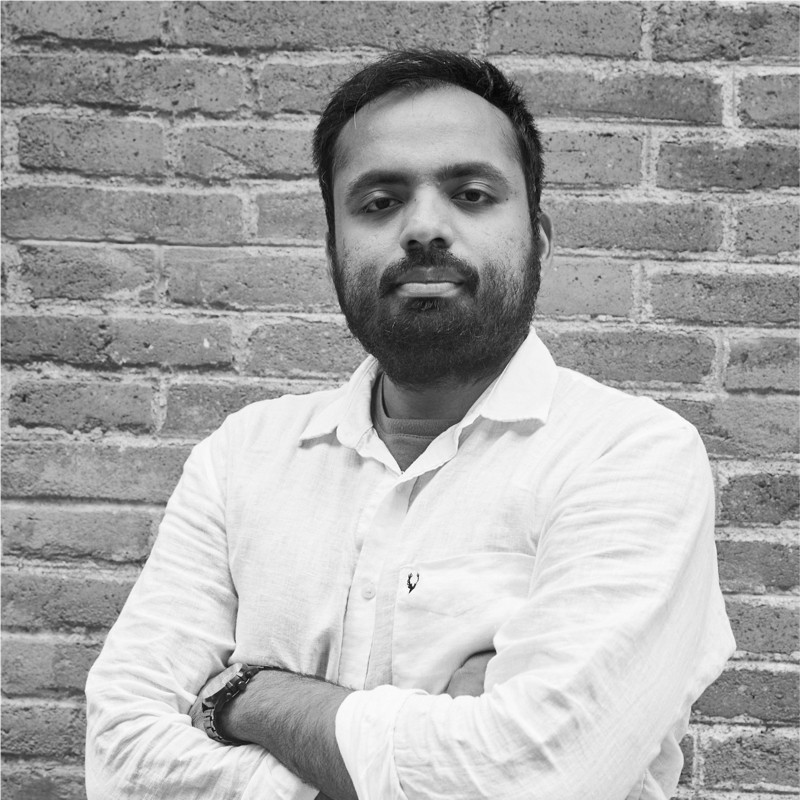}}]{Akshay Jain} received his B.Tech in Electronics and Communication engineering
from NIT-Surat, India (2013), MS in Electrical and Computer Engineering from University of Florida, USA (2015), and PhD in Network Engineering from Universitat Politècnica de Catalunya (2020). Akshay then joined a drone startup Queen B Robotics (2015-2016), where he was also the lead Communications system engineer. During his time as a PhD student, Akshay was a recipient of the prestigious Marie Curie Sklowdowska Actions fellowship from the European Union. In 2011, he also received the prestigious KVPY grant from the Department of Science and Technology, Government of India. He has co-authored more than 20 research papers, which have been published in top tier conference and journals. In November 2019, Akshay joined the NaNoNetworking Center of Catalunya (N3CAT) as a postdoctoral researcher. Subsequently, he joined Neutroon Technologies SL, where he is currently the VP of Telecommunications Engineering department. Here he currently works on the frontiers of wireless communications, optimization methods and machine learning techniques, resource allocation algorithms, Private 5G networks, etc.
\end{IEEEbiography}
\vspace{-1cm}

\begin{IEEEbiography}[{\includegraphics[width=1in,height=1.25in,clip,keepaspectratio]{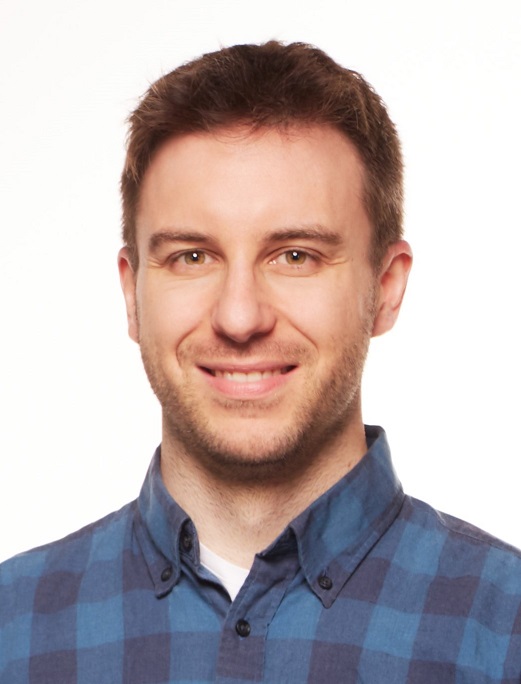}}]{Sergi Abadal} is Project Director at the NaNoNetworking Center in Catalonia, Universitat Politècnica de Catalunya, where he also obtained his PhD in computer science engineering (2016). He has co-authored more than 80 research papers. In 2013, he was awarded by INTEL within his Doctoral Student Honor Program. He is Associate Editor of the Nano Communication Networks (Elsevier) Journal. His research interests include on-chip networking, many-core architectures, and graphene-based wireless communications.
\end{IEEEbiography}
\vspace{-1cm}

\begin{IEEEbiography}[{\includegraphics[width=1in,height=1.25in,clip,keepaspectratio]{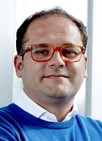}}]{GABRIELE GRADONI} received the Ph.D. degree in electromagnetics from the Universita Politecnica delle Marche, Ancona, Italy, in 2010. He was a Visiting Researcher with the Time, Quantum, and Electromagnetics Team, National Physical Laboratory, Teddington, U.K., in 2008. From 2010 to 2013, he was a Research Associate with the Institute for Research in Electronics and Applied Physics, University of Maryland, College Park, MD, USA. From 2013 to 2016, he was a Research Fellow with the School of Mathematical Sciences, University of Nottingham, U.K. Since 2020, he has been an Associate Professor of mathematics and electrical engineering with the University of Nottingham. His research interests include probabilistic and asymptotic methods for propagation in complex wave systems, wave chaos, and metasurfaces, with applications to electromagnetic compatibility and modern wireless communication systems. He is a member of the American Physical Society, and the Italian Electromagnetics Society. He received the URSI Commission B Young Scientist Award in 2010 and 2016, the Gaetano Latmiral Prize in 2015, and the and the Honorable Mention IEEE TEMC Richard B. Schulz Transactions Prize Paper Award in 2020. From 2014 to 2021, he has been the URSI Commission E Early Career Representative. Since 2020, he has been a Royal Society Industry Fellow at the Maxwell Centre, Cavendish Laboratory, University of Cambridge, U.K., and an Adjunct Associate Professor at the Department of Electrical and Computer Engineering, University of Illinois, Urbana Champaign, U.S.A.
\end{IEEEbiography}
\vspace{-1cm}

\begin{IEEEbiography}[{\includegraphics[width=1in,height=1.25in,clip,keepaspectratio]{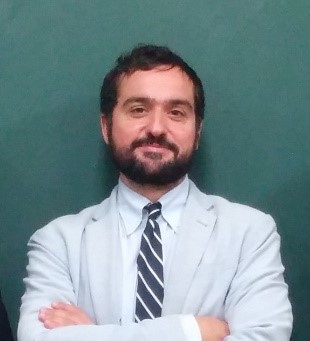}}]{Eduard Alarcón} is an associate professor at the Universitat Politècnica de Catalunya, where he obtained his PhD in electrical engineering in 2000. He has coauthored more than 400 scientific publications, 8 book chapters and 12 patents. He was elected IEEE CAS society distinguished lecturer, member of the IEEE CAS Board of Governors (2010-2013), Associate Editor for IEEE TCAS-I, TCAS-II, JOLPE, and Editor-in-Chief of JETCAS. His research interests include nanocommunications and wireless energy transfer.
\end{IEEEbiography}
\vspace{-1cm}

\begin{IEEEbiography}[{\includegraphics[width=1in,height=1.25in,clip,keepaspectratio]{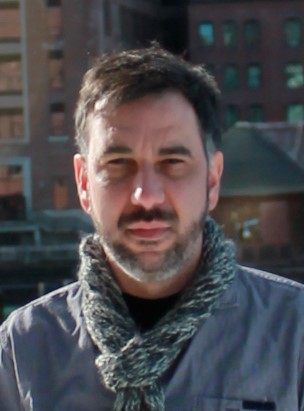}}]{Albert Cabellos-Aparicio} is a full professor at Universitat Politècnica de Catalunya, where he obtained his PhD in computer science engineering in 2008. He is co-founder and scientific director of the NaNoNetworking Center in Catalunya. He has been a visiting researcher at Cisco Systems and Agilent Technologies and a visiting professor at the KTH, Sweden, and the MIT, USA. He has co-authored more than 80 research papers. His research interests include nanocommunications and software-defined networking.
\end{IEEEbiography}

\vfill


\begin{thebibliography}{10}

\providecommand{\url}[1]{#1}
\csname url@samestyle\endcsname
\providecommand{\newblock}{\relax}
\providecommand{\bibinfo}[2]{#2}
\providecommand{\BIBentrySTDinterwordspacing}{\spaceskip=0pt\relax}
\providecommand{\BIBentryALTinterwordstretchfactor}{4}
\providecommand{\BIBentryALTinterwordspacing}{\spaceskip=\fontdimen2\font plus
\BIBentryALTinterwordstretchfactor\fontdimen3\font minus
  \fontdimen4\font\relax}
\providecommand{\BIBforeignlanguage}[2]{{%
\expandafter\ifx\csname l@#1\endcsname\relax
\typeout{** WARNING: IEEEtran.bst: No hyphenation pattern has been}%
\typeout{** loaded for the language `#1'. Using the pattern for}%
\typeout{** the default language instead.}%
\else
\language=\csname l@#1\endcsname
\fi
#2}}
\providecommand{\BIBdecl}{\relax}
\BIBdecl

\bibitem{1309810}
S.~{Cherry}, ``Edholm's law of bandwidth,'' \emph{IEEE Spectrum}, vol.~41,
  no.~7, pp. 58--60, July 2004.

\bibitem{AKYILDIZ20103}
I.~F. Akyildiz and J.~M. Jornet, ``Electromagnetic wireless nanosensor
  networks,'' \emph{Nano Communication Networks}, vol.~1, no.~1, pp. 3 -- 19,
  2010.
  

\bibitem{1491267}
M.~{Marcus} and B.~{Pattan}, ``Millimeter wave propagation: spectrum management
  implications,'' \emph{IEEE Microwave Magazine}, vol.~6, no.~2, pp. 54--62,
  June 2005.

\bibitem{hunukumbure2018mmwave}
M.~Hunukumbure, J.~Luo, M.~Castaneda, R.~DErrico, P.~Zetterberg, A.~A. Zaidi,
  J.~Vihriala, and D.~Giustiniano, ``{Mm-wave specific challenges in designing
  5G transceiver architectures and air-interfaces},'' 2018.

\bibitem{6206517}
L.~{Subrt} and P.~{Pechac}, ``Controlling propagation environments using
  intelligent walls,'' in \emph{2012 6th European Conference on Antennas and
  Propagation (EUCAP)}, 2012, pp. 1--5.

\bibitem{8288263}
S.~{Hu}, F.~{Rusek}, and O.~{Edfors}, ``Cramér-rao lower bounds for
  positioning with large intelligent surfaces,'' in \emph{2017 IEEE 86th
  Vehicular Technology Conference (VTC-Fall)}, 2017, pp. 1--6.
  
  \bibitem{202000783}
O.~Tsilipakos, A.~C.~Tasolamprou, A.~Pitilakis, F.~Liu, X.~Wang, M.~S.~Mirmoosa, D.~C.~Tzarouchis, S.~Abadal, H.~Taghvaee, C.~Liaskos, A.~Tsioliaridou, J.~Georgiou, A.~Cabellos-Aparicio, E.~Alarcón, S.~Ioannidis, A.~Pitsillides, I.~F.~Akyildiz, and N.~V.~Kantartzis, E.~.N.~Economou, C.~M.~Soukoulis, M.~Kafesaki, and S.~Tretyakov,
  ``Toward Intelligent Metasurfaces: The Progress from Globally Tunable Metasurfaces to Software-Defined Metasurfaces with an Embedded Network of Controllers,'' \emph{Advanced Optical Materials}, vol.~8, no.~17, pp. 2000783, 2020.

\bibitem{DAJER2021}
M.~Dajer, Z.~Ma, L.~Piazzi, N.~Prasad, X.-F. Qi, B.~Sheen, J.~Yang, and G.~Yue,
  ``Reconfigurable intelligent surface: Design the channel – a new
  opportunity for future wireless networks,'' \emph{Digital Communications and
  Networks}, 2021.

\bibitem{9424177}
Y.~Liu, X.~Liu, X.~Mu, T.~Hou, J.~Xu, M.~Di~Renzo, and N.~Al-Dhahir,
  ``Reconfigurable intelligent surfaces: Principles and opportunities,''
  \emph{IEEE Communications Surveys Tutorials}, vol.~23, no.~3, pp. 1546--1577,
  2021.

\bibitem{RISE-6G2021}
E.~Calvanese~Strinati, G.~C. Alexandropoulos, V.~Sciancalepore, M.~D. Renzo,
  H.~Wymeersch, D.-T. Phan-huy, M.~Crozzoli, R.~D'Errico, E.~D. Carvalho,
  P.~Popovski, P.~D. Lorenzo, L.~Bastianelli, M.~Belouar, J.~E. Mascolo,
  G.~Gradoni, S.~Phang, G.~Lerosey, and B.~Denis, ``Wireless environment as a
  service enabled by reconfigurable intelligent surfaces: The {RISE-6G}
  perspective,'' \emph{Proc. of EUCNC 6G Summit}, Porto, Portugal, June 2021.

\bibitem{https://doi.org/10.1002/smtd.201600064}
H.-H. Hsiao, C.~H. Chu, and D.~P. Tsai, ``Fundamentals and applications of
  metasurfaces,'' \emph{Small Methods}, vol.~1, no.~4, p. 1600064, 2017.

\bibitem{Li2018}
A.~Li, S.~Singh, and D.~Sievenpiper, ``{Metasurfaces and their applications},''
  \emph{Nanophotonics}, vol.~7, no.~6, pp. 989--1011, 2018.

\bibitem{Chen2016}
H.~T. Chen, A.~J. Taylor, and N.~Yu, ``{A review of metasurfaces: Physics and
  applications},'' \emph{Reports on Progress in Physics}, vol.~79, no.~7, 2016.

\bibitem{Taghvaee2014}
H.~Taghvaee, M.~Seyyedi, and A.~Rezaee, ``{Design of a Metamaterial Dual Band
  Absorber},'' in \emph{The third Iranian Conference on Engineering
  Electromagnetic ICEEM}, vol.~3.\hskip 1em plus 0.5em minus 0.4em\relax
  Tehran: Iranian Scientific Society of Engineering Electromagnetics, 2014.

\bibitem{Taghvaee2017a}
H.~R. Taghvaee, H.~Nasari, and M.~S. Abrishamian, ``{Circuit modeling of
  graphene absorber in terahertz band},'' \emph{Optics Communications}, vol.
  383, pp. 11--16, 2017.

\bibitem{Arbabi2017}
A.~Arbabi, E.~Arbabi, Y.~Horie, S.~M. Kamali, and A.~Faraon, ``{Planar
  metasurface retroreflector},'' \emph{Nature Photonics}, vol.~11, no.~7, pp.
  415--420, 2017.

\bibitem{Liu2018a}
S.~Liu, P.~P. Vabishchevich, A.~Vaskin, J.~L. Reno, G.~A. Keeler, M.~B.
  Sinclair, I.~Staude, and I.~Brener, ``{An all-dielectric metasurface as a
  broadband optical frequency mixer},'' \emph{Nature Communications}, vol.~9,
  no.~1, pp. 1--6, 2018.
  
  \bibitem{PhysRevB.104.235409}
H.~Taghvaee, F.~Zarrinkhat, F.~Liu, A.~D\'{\i}az-Rubio, and S.~Tretyakov, ``{Subwavelength focusing by engineered power-flow conformal metamirrors},'' \emph{Phys. Rev. B}, vol.~104,
  no.~23, pp. 235409, 2021.

\bibitem{Taghvaee2017}
H.~R. Taghvaee, F.~Zarrinkhat, and M.~S. Abrishamian, ``{Terahertz Kerr
  nonlinearity analysis of a microribbon graphene array using the harmonic
  balance method},'' \emph{Journal of Physics D: Applied Physics}, vol.~50,
  no.~25, 2017.

\bibitem{doi:10.1002/mop.32164}
L.~Bao and T.~J. Cui, ``Tunable, reconfigurable, and programmable
  metamaterials,'' \emph{Microwave and Optical Technology Letters}, vol.~62,
  no.~1, pp. 9--32, 2020.
  
  
  \bibitem{s21082765}
H.~Taghvaee, A.~Jain, X.~Timoneda, C.~Liaskos, S.~Abadal, E.~Alarón, and A.~Cabellos-Aparicio, ``{Radiation Pattern Prediction for Metasurfaces: A Neural Network-Based Approach},'' \emph{Sensors}, vol.~21, no.~8, pp. 2765, 2021.

\bibitem{Ma2019}
Q.~Ma, G.~Bai, H.~Jing \emph{et~al.}, ``{Smart metasurface with self-adaptively
  reprogrammable functions},'' \emph{Light: Science {\&} Applications}, 2019.

\bibitem{8788546}
A.~C.~Tasolamprou, A.~Pitilakis, S.~Abadal, O.~Tsilipakos, X.~Timoneda,
  H.~Taghvaee, M.~Sajjad~Mirmoosa, F.~Liu, C.~Liaskos, A.~Tsioliaridou,
  S.~Ioannidis, N.~V. Kantartzis, D.~Manessis, J.~Georgiou,
  A.~Cabellos-Aparicio, E.~Alarcón, A.~Pitsillides, I.~F. Akyildiz, S.~A.
  Tretyakov, E.~N. Economou, M.~Kafesaki, and C.~M. Soukoulis, ``Exploration of
  intercell wireless millimeter-wave communication in the landscape of
  intelligent metasurfaces,'' \emph{IEEE Access}, vol.~7, pp.
  122\,931--122\,948, 2019.
  
  \bibitem{9737695}
H.~Taghvaee, A.~Pitilakis, O.~Tsilipakos, A.~C.~Tasolamprou, N.~v.~Kantartzis, M.~Kafesaki, A.~Cabellos-Aparicio, E.~Alarcon, and S.~Abadal, ``Multiwideband Terahertz Communications Via Tunable Graphene-Based Metasurfaces in 6G Networks: Graphene Enables Ultimate Multiwideband THz Wavefront Control,'' \emph{IEEE Vehicular Technology Magazine}, vol.~17, no.~2, pp. 16 -- 25, 2022.

\bibitem{PhysRevApplied.11.044024}
F.~Liu, O.~Tsilipakos, A.~Pitilakis, A.~C. Tasolamprou, M.~S. Mirmoosa, N.~V.
  Kantartzis, D.-H. Kwon, J.~Georgiou, K.~Kossifos, M.~A. Antoniades,
  M.~Kafesaki, C.~M. Soukoulis, and S.~A. Tretyakov, ``Intelligent metasurfaces
  with continuously tunable local surface impedance for multiple reconfigurable
  functions,'' \emph{Phys. Rev. Applied}, vol.~11, p. 044024, Apr 2019.

\bibitem{8745693}
S.~E. Hosseininejad, K.~Rouhi, M.~Neshat, A.~Cabellos-Aparicio, S.~Abadal, and
  E.~Alarcón, ``Digital metasurface based on graphene: An application to beam
  steering in terahertz plasmonic antennas,'' \emph{IEEE Transactions on
  Nanotechnology}, vol.~18, pp. 734--746, 2019.

\bibitem{V2X}
Y.~U. Ozcan, O.~Ozdemir, and G.~K. Kurt, ``Reconfigurable intelligent surfaces
  for the connectivity of autonomous vehicles,'' \emph{IEEE Transactions on
  Vehicular Technology}, vol.~70, no.~3, pp. 2508--2513, 2021.
  
  \bibitem{9593176}
T.~Saeed, W.~Aziz, A.~Pitsillides, V.~Vassiliou, I.~.F.~Akyildiz, H.~Taghvaee, S.~Abadal, C.~Liaskos, A.~Tsioliaridou, S.~Ioannidis, E.~Emoyon-Iredia, M.~Lestas, ``On the Use of Programmable Metasurfaces in Vehicular Networks,'' \emph{IEEE 22nd International Workshop on Signal Processing Advances in Wireless Communications (SPAWC)}, pp. 521--525, 2021.

\bibitem{9171580}
A.~Pitilakis, O.~Tsilipakos, F.~Liu, K.~M. Kossifos, A.~C. Tasolamprou, D.-H.
  Kwon, M.~S. Mirmoosa, D.~Manessis, N.~V. Kantartzis, C.~Liaskos, M.~A.
  Antoniades, J.~Georgiou, C.~M. Soukoulis, M.~Kafesaki, and S.~A. Tretyakov,
  ``A multi-functional reconfigurable metasurface: Electromagnetic design
  accounting for fabrication aspects,'' \emph{IEEE Transactions on Antennas and
  Propagation}, vol.~69, no.~3, pp. 1440--1454, 2021.

\bibitem{8753713}
L.~Bao, R.~Y. Wu, X.~Fu, Q.~Ma, G.~D. Bai, J.~Mu, R.~Jiang, and T.~J. Cui,
  ``Multi-beam forming and controls by metasurface with phase and amplitude
  modulations,'' \emph{IEEE Transactions on Antennas and Propagation}, vol.~67,
  no.~10, pp. 6680--6685, 2019.
  
  \bibitem{Karimipour2019}
M.~Karimipour, N.~Komjani, and I.~Aryanian,
  ``Shaping Electromagnetic Waves with Flexible and Continuous Control of the Beam Directions Using Holography and Convolution Theorem,'' \emph{Scientific Reports}, vol.~9,
  no.~1, pp. 11825, 2019.

  \bibitem{Taghvaee_2021}
H.~Taghvaee,
  ``On scalable, reconfigurable, and intelligent metasurfaces,'' Doctoral thesis, UPC, Department of Computer Architecture, 2021.
  

  \bibitem{9769001}
H.~Taghvaee, S.~Terranova, N.~M.~Mohammed, and G.~Gradoni,
  ``Sustainable Multi-User Communication with Reconfigurable Intelligent Surfaces in 5G Wireless Networks and Beyond,'' 16th European Conference on Antennas and Propagation (EuCAP), 2022.
  
\bibitem{DingChen}
G.~Ding, K.~Chen, X.~Luo, G.~Qian, J.~Zhao, T.~Jiang, and Y.~Feng, ``Direct
  routing of intensity-editable multi-beams by dual geometric phase
  interference in metasurface,'' \emph{Nanophotonics}, vol.~9, no.~9, pp.
  2977--2987, 2020.

\bibitem{Froehlich1991}
E.~Froehlich and A.~Kent, \emph{{The Froehlich/Kent encyclopedia of
  telecommunications: Volume 2}}, 1st~ed.\hskip 1em plus 0.5em minus
  0.4em\relax CRC Press, 1991.

\bibitem{9220095}
H.~B. {Sedeh}, M.~M. {Salary}, and H.~{Mosallaei}, ``Adaptive multichannel
  terahertz communication by space-time shared aperture metasurfaces,''
  \emph{IEEE Access}, pp. 1--1, 2020.

\bibitem{7504504}
K.~I. Pedersen, F.~Frederiksen, G.~Berardinelli, and P.~E. Mogensen, ``The
  coverage-latency-capacity dilemma for tdd wide area operation and related 5g
  solutions,'' in \emph{2016 IEEE 83rd Vehicular Technology Conference (VTC
  Spring)}, 2016, pp. 1--5.

\bibitem{7945855}
S.-Y. Lien, S.-L. Shieh, Y.~Huang, B.~Su, Y.-L. Hsu, and H.-Y. Wei, ``5g new
  radio: Waveform, frame structure, multiple access, and initial access,''
  \emph{IEEE Communications Magazine}, vol.~55, no.~6, pp. 64--71, 2017.
  
\bibitem{9306896}
N. Marzieh, J. Vahid, S. Robert, P. H. Vincent, ``Physics-Based Modeling and Scalable Optimization of Large Intelligent Reflecting Surfaces,'' \emph{IEEE Transactions on Communications}, vol.~69, no.~4, pp. 2673--2691, 2021.

\bibitem{CST}
\BIBentryALTinterwordspacing
``{CST Microwave Studio}.'' [Online]. Available: \url{http://www.cst.com}
\BIBentrySTDinterwordspacing

\bibitem{ashoor2020metasurface}
A.~Z. Ashoor and S.~Gupta, ``Metasurface reflector with real-time independent
  magnitude and phase control,'' 2020.

\bibitem{Hashemi2016ABM}
M.~Hashemi, A.~Moazami, M.~Naserpour, and C.~J. Zapata-Rodriguez, ``A broadband
  multifocal metalens in the terahertz frequency range,'' \emph{Optics
  Communications}, vol. 370, pp. 306--310, 2016.

\bibitem{Wang2018}
X.~Wang, J.~Ding, B.~Zheng \emph{et~al.}, ``Simultaneous realization of
  anomalous reflection and transmission at two frequencies using bi-functional
  metasurfaces,'' \emph{Scientific Report}, vol.~8, p. 1876, 2018.

\bibitem{9109701}
H.~Taghvaee, S.~Abadal, A.~Pitilakis, O.~Tsilipakos, A.~C. Tasolamprou,
  C.~Liaskos, M.~Kafesaki, N.~V. Kantartzis, A.~Cabellos-Aparicio, and
  E.~Alarcón, ``Scalability analysis of programmable metasurfaces for beam
  steering,'' \emph{IEEE Access}, vol.~8, pp. 105320--105334, 2020.
 
\bibitem{9350282}
V.~Jamali, M.~Najafi, R.~Schober, and H.~Vincent Poor, ``Power Efficiency, Overhead, and Complexity Tradeoff of IRS Codebook Design—Quadratic Phase-Shift Profile,'' \emph{IEEE Communications Letters}, vol.~25, no.~6, pp. 2048--2052, 2021.

  
\bibitem{Wang2014}
K.~Wang, J.~Zhao, Q.~Cheng \emph{et~al.}, ``Broadband and broad-angle
  low-scattering metasurface based on hybrid optimization algorithm,''
  \emph{Scientific Report}, vol.~4, p. 5935, 2014.

\bibitem{Report2018b}
3GPP, ``{TR 138 901 - V15.0.0 - 5G; Study on channel model for frequencies from
  0.5 to 100 GHz (3GPP TR 38.901 version 15.0.0 Release 15)},'' 2018.

\bibitem{9110915}
X.~Hu, C.~Zhong, Y.~Zhu, X.~Chen, and Z.~Zhang, ``Programmable
  metasurface-based multicast systems: Design and analysis,'' \emph{IEEE
  Journal on Selected Areas in Communications}, vol.~38, no.~8, pp. 1763--1776,
  2020.

\bibitem{8796365}
E.~Basar, M.~Di~Renzo, J.~De~Rosny, M.~Debbah, M.-S. Alouini, and R.~Zhang,
  ``Wireless communications through reconfigurable intelligent surfaces,''
  \emph{IEEE Access}, vol.~7, pp. 116\,753--116\,773, 2019.

\bibitem{9541182}
E.~Basar, I.~Yildirim, and F.~Kilinc, ``Indoor and outdoor physical channel
  modeling and efficient positioning for reconfigurable intelligent surfaces in
  mmwave bands,'' \emph{IEEE Transactions on Communications}, vol.~69, no.~12,
  pp. 8600--8611, 2021.

\bibitem{Ozdogan2020c}
O.~Ozdogan, E.~Bjornson, and E.~G. Larsson, ``{Intelligent Reflecting Surfaces:
  Physics, Propagation, and Pathloss Modeling},'' \emph{IEEE Wirel. Commun.
  Lett.}, vol.~9, no.~5, pp. 581--585, 2020.

\bibitem{Bjornson2020b}
E.~Bjornson, O.~Ozdogan, and E.~G. Larsson, ``{Intelligent Reflecting Surface
  Versus Decode-and-Forward: How Large Surfaces are Needed to Beat Relaying?}''
  \emph{IEEE Wirel. Commun. Lett.}, vol.~9, no.~2, pp. 244--248, 2020.

\bibitem{jain2020user}
A.~Jain, E.~Lopez-Aguilera, and I.~Demirkol, ``{User Association and Resource
  Allocation in 5G (AURA-5G): A Joint Optimization Framework},'' 2020.

\bibitem{jain2020mobility}
A.~\hspace{0mm}Jain, E.~Lopez-Aguilera, and I.~Demirkol, ``{Are mobility
  management solutions ready for 5G and beyond?}'' \emph{Computer
  Communications}, vol. 161, pp. 50--75, Sept. 2020.

\bibitem{MichaMaterniaNokia2016}
{Micha{\l} Maternia} and S.~E.~E. Ayoubi, ``{5G PPP use cases and performance
  evaluation models},'' \emph{5G-PPP Initiat.}, 2016.

\bibitem{Specification2018}
T.~Specification, ``{System Architecture for the 5G System (3GPP TS 23.501
  version 15.2.0 Release 15)},'' 2018.

\bibitem{8386686}
S.~Sun, T.~S. Rappaport, M.~Shafi, P.~Tang, J.~Zhang, and P.~J. Smith,
  ``Propagation models and performance evaluation for 5g millimeter-wave
  bands,'' \emph{IEEE Transactions on Vehicular Technology}, vol.~67, no.~9,
  pp. 8422--8439, 2018.

\end{thebibliography}
\end{document}